\documentclass[longauth]{aa}  

\usepackage{subcaption}
\usepackage{graphicx}
\usepackage{comment}
\usepackage[dvipsnames]{xcolor}
\usepackage{multirow}
\usepackage{hhline}
\usepackage{ulem}
\usepackage{amsmath}
\usepackage{amssymb}
\usepackage{txfonts}
\usepackage{hyperref}
\hypersetup{colorlinks,linkcolor={blue},citecolor={blue}}

\newcommand{\planck}{{\em Planck\,}}
\newcommand{\xmm}{{\em XMM-Newton\,}}

\def\xmm{{\it XMM-Newton}}
\def\planck{{\it Planck}}

\begin{document}

   \title{CHEX-MATE: towards a consistent universal pressure profile and cluster mass reconstruction
   }
   \titlerunning{CHEX-MATE: towards a consistent universal pressure profile and cluster mass reconstruction}

  \author{
  M.~Muñoz-Echeverr\'ia\inst{\ref{IRAP}}\thanks{miren.munoz-echeverria@irap.omp.eu}, 
  E.~Pointecouteau \inst{\ref{IRAP}}, 
  G.~W.~Pratt \inst{\ref{CEA}}, 
  J.-F.~Mac\'ias-P\'erez\inst{\ref{LPSC}}, 
  M.~Douspis \inst{\ref{IAS}}, 
  L.~Salvati \inst{\ref{IAS}}, 
  I.~Bartalucci  \inst{\ref{inaf-mi}},
  H.~Bourdin \inst{\ref{torvergata1}, \ref{torvergata2}}, 
  N.~Clerc \inst{\ref{IRAP}}, 
  F.~De~Luca \inst{\ref{torvergata1}, \ref{torvergata2}}, 
  M.~De~Petris \inst{\ref{sapienza}},
  M.~Donahue \inst{\ref{msu}},
  S.~Dupourqu\'e \inst{\ref{IRAP}}, 
  D.~Eckert \inst{\ref{uni_geneva}},
  S.~Ettori \inst{\ref{inaf-oas},\ref{infn-bo}},
  M.~Gaspari \inst{\ref{uni-modena}},
  F.~Gastaldello \inst{\ref{inaf-mi}},
  M.~Gitti \inst{\ref{uni-bo}, \ref{ira}},
  A.~Gorce \inst{\ref{IAS}}, 
  S.~Ili\'c \inst{\ref{ijclab}},
  S.~T.~Kay  \inst{\ref{manchester}}, 
  J.~Kim \inst{\ref{kaist}}, 
  L.~Lovisari \inst{\ref{inaf-mi}, \ref{cfa}},
  B.~J.~Maughan \inst{\ref{bristol}},
  P.~Mazzotta \inst{\ref{torvergata1}, \ref{torvergata2}},
  L.~McBride \inst{\ref{IAS}},
  J.-B.~Melin  \inst{\ref{IRFU}},
  F.~Oppizzi \inst{\ref{infn-genova}}
  E.~Rasia \inst{\ref{inaf_trieste},\ref{ifpu_trieste},\ref{michigan}}, 
  M.~Rossetti \inst{\ref{inaf-mi}},
  H.~Saxena \inst{\ref{caltech}},
  J.~Sayers \inst{\ref{caltech}}, 
  M.~Sereno \inst{\ref{inaf-oas},\ref{infn-bo}},
  M.~Tristram \inst{\ref{ijclab}}
  }

 \institute{
      IRAP, CNRS, Université de Toulouse, CNES, UT3-UPS, (Toulouse), France
     \label{IRAP}
     \and
     Universit\'e Paris-Saclay, Universit\'e Paris Cit\'e, CEA, CNRS, AIM, 91191, Gif-sur-Yvette, France
     \label{CEA}
     \and
     Université Grenoble Alpes, CNRS, LPSC-IN2P3, 53, avenue
    des Martyrs, 38000 Grenoble, France
     \label{LPSC} 
     \and
     Université Paris-Saclay, CNRS, Institut d’Astrophysique Spatiale, 91405, Orsay, France
     \label{IAS}
    \and
    INAF, Istituto di Astrofisica Spaziale e Fisica Cosmica di Milano, via A. Corti 12, 20133 Milano, Italy \label{inaf-mi}
    \and
    Dipartimento di Fisica, Universit\`a degli studi di Roma Tor Vergata, Via della Ricerca Scientifica 1, I-00133 Roma, Italy \label{torvergata1}
    \and
    INFN, Sezione di Roma ‘Tor Vergata’, Via della Ricerca Scientifica, 1, 00133, Roma, Italy \label{torvergata2}
    \and
    Dipartimento di Fisica, Sapienza Università di Roma, Piazzale Aldo Moro 5, I-00185 Rome, Italy \label{sapienza}
    \and
    Michigan State University (MSU) - Dept of Physics and Astronomy, 567 Wilson Road, East Lansing, Michigan, 48824, USA \label{msu}
    \and
    Department of Astronomy, University of Geneva, Ch. d’Ecogia 16, CH-1290 Versoix, Switzerland \label{uni_geneva}     
    \and
     INAF - Osservatorio di Astrofisica e Scienza dello Spazio di Bologna, via Piero Gobetti 93/3, I-40129 Bologna, Italy \label{inaf-oas}
    \and
    INFN, Sezione di Bologna, viale Berti Pichat 6/2, I-40127 Bologna, Italy \label{infn-bo}
    \and
    Department of Physics, Informatics and Mathematics, University of Modena and Reggio Emilia, 41125 Modena, Italy \label{uni-modena}
    \and
    Dipartimento di Fisica e Astronomia (DIFA), Alma Mater Studiorum - Università di Bologna, via Gobetti 93/2, 40129 Bologna, Italy \label{uni-bo}
    \and
    Istituto Nazionale di Astrofisica – Istituto di Radioastronomia (IRA), via Gobetti 101, 40129 Bologna, Italy \label{ira}
    \and
    IJCLab, Université Paris-Saclay, CNRS/IN2P3, IJCLab, 91405 Orsay, France \label{ijclab}
    \and
    Jodrell Bank Centre for Astrophysics, Department of Physics and Astronomy, The University of Manchester, Manchester M13 9PL, UK  \label{manchester}
    \and
    Department of Physics, Korea Advanced Institute of Science and Technology (KAIST), 291 Daehak-ro, Yuseong-gu, Daejeon 34141, Republic of Korea \label{kaist}    
    \and
    Center for Astrophysics $|$ Harvard $\&$ Smithsonian, 60 Garden Street, Cambridge, MA 02138, USA \label{cfa}    
    \and
    HH Wills Physics Laboratory, University of Bristol, Bristol, UK \label{bristol}
    \and
    IRFU, CEA, Universit\'e Paris-Saclay, 91191, Gif-sur-Yvette, France
    \label{IRFU}
    \and
    INFN-Sezione di Genova, Via Dodecaneso 33, 16146, Genova, Italy \label{infn-genova}
    \and
    INAF, Osservatorio Astronomico di Trieste, via Tiepolo 11, I-34131, Trieste, Italy \label{inaf_trieste}
    \and
    IFPU, Institute for Fundamental Physics of the Universe, Via Beirut 2, 34014 Trieste, Italy \label{ifpu_trieste}
    \and
    Department of Physics; University of Michigan, Ann Arbor, MI 48109, USA \label{michigan}
    \and
    California Institute of Technology, 1200 East California Boulevard, Pasadena, California, USA \label{caltech}
   }

   \date{Received --------; accepted -------}

  \abstract
  {
  In a self-similar paradigm of structure formation, the thermal pressure of the hot intra-cluster gas follows a universal distribution once the profile of each cluster is normalised based on
  the proper mass and redshift dependencies. The reconstruction of such a universal pressure profile requires an individual estimate of the mass of each cluster. In this context, we present a method to jointly fit, for the first time, the universal pressure profile and individual cluster $M_{500}$ masses over a sample of galaxy clusters, properly accounting for correlations between the profile shape and amplitude, and masses scaling the individual profiles. We demonstrate the power of the method and show that a consistent exploitation of the universal pressure profile and cluster mass estimates when modelling the thermal pressure in clusters is necessary to avoid biases. In particular, the method, informed by a cluster mass scale, outputs individual cluster masses with same accuracy and better precision than input masses.
  Using data from the
  «Cluster HEritage project with \xmm\ Mass Assembly and Thermodynamics at the Endpoint of structure formation», we investigate a sample of $\sim 25$ galaxy clusters spanning mass and redshift ranges of $2 \lesssim M_{500}/10^{14} \; \mathrm{M}_{\odot} \lesssim 14$ and $0.07 < z < 0.6$.
    }


   \authorrunning{M.~Muñoz-Echeverr\'ia et al.}
   \maketitle

\section{Introduction}
\label{sec:intro}

Galaxy clusters are formed by the accretion of matter at the nodes of the cosmic web. If a cluster is virialised and its intra-cluster medium (ICM) fully thermalised, the thermal pressure counter balances the effect of gravitational forces. Since gravity does not have preferred scales, clusters are nearly scaled versions of
each other \citep[see, e.g., the reviews in][]{2002ARA&A..40..539R,2005RvMP...77..207V}, and consequently, the thermal pressure distribution in galaxy clusters is universal (in the intermediate radial range) when normalised with respect to mass and redshift \citep[][hereafter N07 and A10]{2007ApJ...668....1N,2010AA...517A..92A}.

Galaxy cluster detection algorithms for millimetre wavelength observations \citep{2006A&A...459..341M,2009MNRAS.393..681C} and cosmological analyses based on the thermal Sunyaev-Zel'dovich (tSZ) angular power spectrum \citep{2017MNRAS.469..394H,2018A&A...614A..13S,2018MNRAS.477.4957B} rely on analytical universal pressure profile (UPP) functionals, calibrated beforehand with dedicated observations. The impact on cosmological results of the assumed profile has been shown to be non-negligible, with different choices shifting the cosmological results, both for tSZ angular power spectrum and cluster number count analyses \citep{2019MNRAS.490..784R, 2022MNRAS.509..300T, 2025ApJ...981..170H, 2024A&A...686A..15G}.

Several studies have characterised the UPP from tSZ and/or X-ray data \citep{2007ApJ...668....1N, 2010AA...517A..92A,2013AA...550A.131P,2013ApJ...768..177S,2014ApJ...794...67M,2016ApJ...832...26S, 2017ApJ...838...86R, 2017ApJ...843...72B, 2018MNRAS.474.1065S, 2019AA...621A..41G, 2021AA...651A..73P, 2021ApJ...908...91H, 2023AA...678A.197M, 2023ApJ...944..221S}. These various works differ in the employed methods, data sets, and cluster samples, but they all parametrise the thermal pressure distribution following the generalised Navarro-Frenk-White (gNFW) model. In the present work, we also consider the gNFW parametrisation. This model was adopted by \cite{2007ApJ...668....1N} \citep[and previously suggested by][]{1996MNRAS.278..488Z} based on galaxy cluster observations and simulations.
The previously-cited studies demonstrated that overall the pressure distribution in galaxy clusters is close to self-similar at intermediate radii where gravity is driving the physical processes. In the core and outskirts, various processes such as accretion shocks, turbulence, or AGN feedback \citep{2020NatAs...4...10G} introduce deviations to the universal pressure model.

The reconstruction of a UPP over a population (sample) of clusters requires the scaling of individual profiles by a characteristic mass scale \citep{1986MNRAS.222..323K}, which in the following we choose to be $M_{500}$,\footnote{\label{M500def}$M_{500}$ and $R_{500}$ are defined as $M_{500} = (4/3) \pi R_{500}^3 \times  500\rho_{\mathrm{crit}}$ where $\rho_{\mathrm{crit}} (z) =3H(z)^{2}/(8\pi G)$ is the critical density of the Universe at the cluster redshift $z$. Similarly, we define $\theta_{500}$ as $\theta_{500} = R_{500}/D_{\mathrm{A}}$, with $D_{\mathrm{A}}$ the angular-diameter distance of the cluster.} following previous studies. This is the mass of a cluster within a sphere of radius $R_{500}^{\ref{M500def}}$.
Thus, the shape and cluster-to-cluster scatter of the derived UPP depends on the manner in which the value of $M_{500}$ is determined \citep[e.g., see figure 7 in][]{2010AA...517A..92A}. Ideally, the true value of $M_{500}$ would be utilised, but direct measurements of this true value are not possible from observations. As a result, most studies rely on observational mass proxies, preferably those with minimal scatter and bias.
Traditionally, hydrostatic equilibrium (HSE) or cluster weak lensing mass estimates have been employed to scale the individual pressure profiles, which are then used to measure the UPP. HSE and lensing mass estimates are known to be biased and scattered, respectively, with respect to the true masses \citep[see][for a review]{pratt2019}. Usually, individual pressure profiles are scaled according to the value of the chosen characteristic mass proxy, and most previous analyses did not propagate uncertainties and biases associated with those mass proxies, leading to potential biases in the inferred UPP. Recently, \cite{2021ApJ...908...91H} studied the impact of the HSE mass bias on the UPP, while the uncertainties of mass estimates have been propagated to the UPP error budget in \cite{2023ApJ...944..221S} (hereafter S23).

As we will show in this paper, it appears that characteristic mass scales and universal profile shape and amplitude (for the pressure in our case) are strongly correlated. They, therefore, need to be jointly extracted from a given set of observations to properly propagate the uncertainties from the mass proxies into the universal profile. Otherwise, cosmological analyses based on the UPP will underestimate the current uncertainties. At the same time, if a consistent propagation of the bias associated with the masses used to scale the profiles is not performed, the tSZ modelling using the UPP, and subsequently inferred cosmological parameters, will be biased. In the following, we present for the first time a method to jointly fit the UPP and minimally biased individual cluster masses.

We present this work in the framework of the ``Cluster HEritage project with \xmm\ – Mass Assembly and Thermodynamics at the Endpoint of structure formation''  \citep[CHEX-MATE\footnote{\url{http://xmm-heritage.oas.inaf.it/}},][]{2021A&A...650A.104C}. The CHEX-MATE project is a Multi-Year  \xmm\ Heritage programme that aims at understanding the interplay of gravitational
and non-gravitational processes in galaxy clusters and their impact on cluster mass estimations. The full sample consists of 118 \planck\ tSZ-selected clusters spanning mass and redshift ranges of $2 \lesssim M_{500}/10^{14} \; \mathrm{M}_{\odot} \lesssim 14$ and $0.05 < z < 0.6$. This work is focused on a subsample of CHEX-MATE clusters for which data is already available (Sect.~\ref{sec:dataset}). In addition to \xmm\ data, the CHEX-MATE project has multi-wavelength coverage of the sample, including radio, millimetre, optical, and infrared data \citep{2021A&A...650A.104C}. In this paper, we make use of X-ray and millimetre wavelength observations to extract thermal pressure profiles (Sect.~\ref{sec:pressprof}), and different cluster mass estimates (Sect.~\ref{sec:m500s}).

The paper is structured as follows. The data set is presented in Sect.~\ref{sec:dataset}. Section~\ref{sec:model} describes the thermal pressure model and the implementation of the fitting method is detailed in Sect.~\ref{sec:fittingmethod}. The results from the application of the method to data are given in Sect.~\ref{sec:data}.
Finally, conclusions are drawn in Sect.~\ref{sec:conclusion}. We assume a flat $\Lambda$CDM cosmology with $H_0 = 70$ km s$^{-1}$ Mpc$^{-1}$ and $\Omega_{\mathrm{m}} = 0.3$. Throughout the paper $G$ corresponds to the Newtonian constant of gravitation, $H(z)$ is the Hubble parameter $H(z) = H_0 \sqrt{\Omega_{\mathrm{m}}(1+z)^3+\Omega_{\Lambda}}$, and $h_{70}= H_0/(70\; \mathrm{km/s/Mpc})$. The logarithm with base 10 is `$\log$' and `$\ln$' refers to the natural logarithm.

\section{Data set}
\label{sec:dataset}

For the present work, we focus on the ``Data Release 1'' (DR1) subsample of CHEX-MATE clusters presented in \cite{2024A&A...686A..68R}. Cluster names, redshifts, and MMF3 masses (see Sect.~\ref{sec:m500s}) are summarised in Table 1 in \cite{2024A&A...686A..68R}. The DR1 clusters have been selected to be representative of the full CHEX-MATE sample in terms of mass, redshift, and X-ray morphology. In this section, we present the data used in this work: the individual pressure profiles for the 28 DR1 clusters\footnote{We excluded PSZ2~G042.81+56.61 and PSZ2~G057.78+52.32 for which \xmm\ observations include off-set pointings not implemented in the current version of the analysis.} and their $M_{500}$ mass estimates.

\subsection{Pressure profiles}
\label{sec:pressprof}

For each cluster, we combine the thermal pressure profiles reconstructed independently from \xmm\ X-ray data and \planck\ tSZ data. Such a combination allows us to cover a large radial range on the individual profiles: the high angular resolution of \xmm\ enables tracing the core (down to $\sim 0.03 \times R_{500}$) and the outskirts are mapped by \planck. In the following, we briefly present the 
reconstruction of the individual thermal pressure profiles. We note that other approaches can be taken to reconstruct the pressure distribution. For instance, in the CHEX-MATE paper by De Luca et al. in prep., a pressure model is fitted jointly to \xmm\ and \planck\ observables following the methodology presented in \cite{2017ApJ...843...72B} (hereafter B17). However, as shown in Appendix B in \cite{2019AA...621A..41G} (hereafter G19), both approaches lead to compatible results. In \cite{2024A&A...686A..97K}, another pressure reconstruction method is introduced: a forward-modelling technique that jointly fits X-ray and tSZ data to infer three dimensional intracluster gas properties.

\begin{figure}
    \centering
    \includegraphics[scale=0.42, trim={1cm 0.5cm 1cm 0.5cm}]{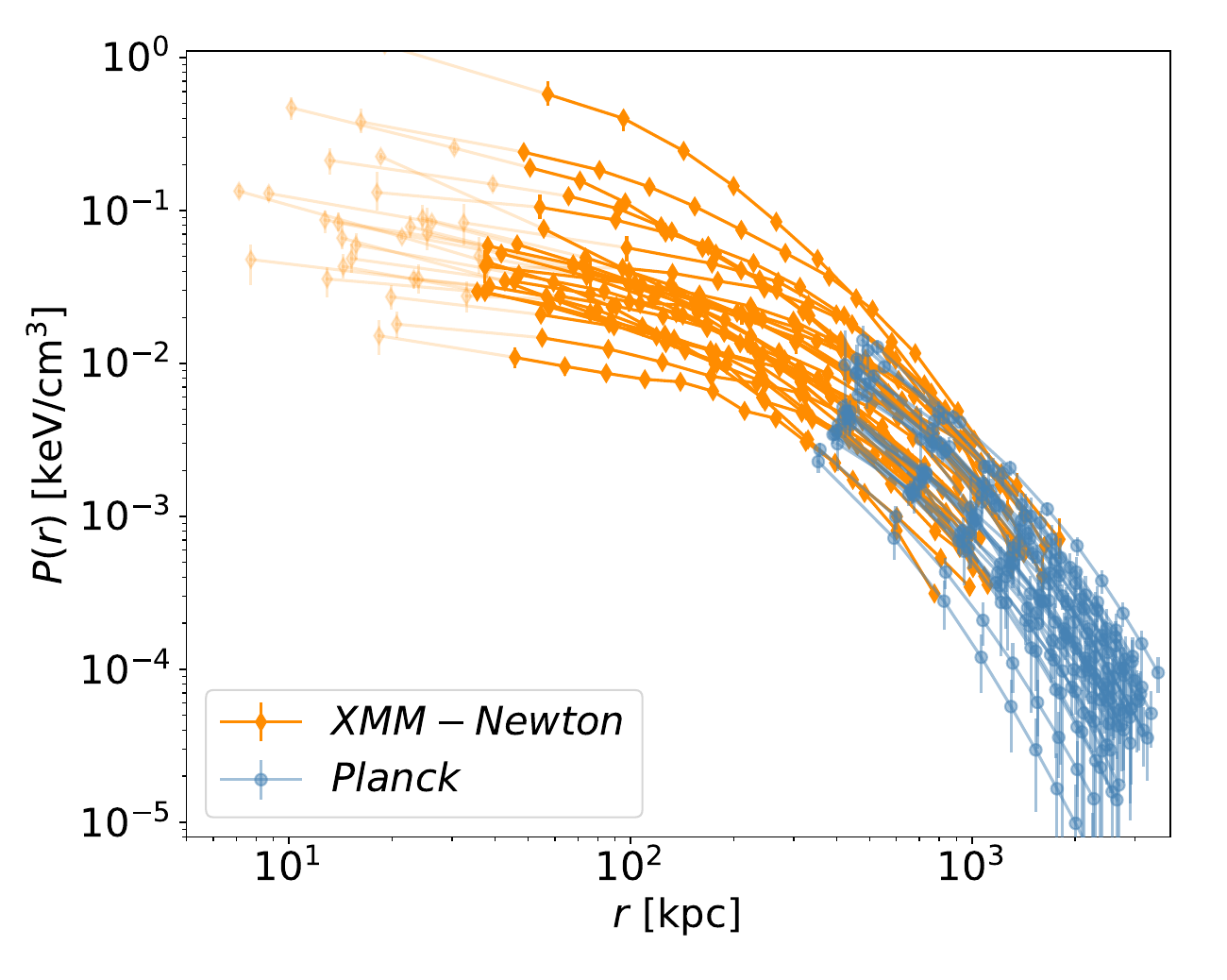}
    \caption{Pressure profiles for the 28 galaxy clusters investigated in this work reconstructed from \xmm\ (orange) and \planck\ (blue) data. \xmm\ data profiles below $r/R_{500}=0.03$ are shaded and these pressure bins are not considered in our analysis (see text).
    }
    \label{fig:pressureprofs}
\end{figure}

\subsubsection{\xmm}
\label{sec:xmm}

The reduction of the raw \xmm\ data was performed using the CHEX-MATE pipeline described in previous works \citep{2023A&A...674A.179B,2024A&A...682A..45L,2024A&A...686A..68R,2024A&A...691A.340R}.
First, data were cleaned by removing the flare events and point sources \citep[see][and references therein]{2023A&A...674A.179B}.
Then, centred on the X-ray peaks, projected surface brightness profiles were reconstructed following the steps detailed in \cite{2023A&A...674A.179B}. For this paper, we considered the azimuthal median surface brightness profiles, and they were converted into emission measure profiles by accounting for the redshift of each cluster and the emissivity in the $0.7-1.2$~keV band \citep{2002A&A...389....1A,2023A&A...674A.179B}. Also centred on the X-ray peaks, the projected temperature profiles were extracted following the spectral fitting method presented in \cite{2024A&A...686A..68R}. 

The deprojection of the profiles was performed as summarised in \cite{2024A&A...691A.340R}. The vignetting-corrected and background-subtracted emission profiles were deprojected into density profiles with the geometrical deprojection and point spread function (PSF) correction methods from \cite{2006A&A...459.1007C}, accounting for the temperature and abundances in each cluster \citep{2003A&A...408....1P}. The binning of surface brightness profiles guarantees a constant signal-to-noise \citep[see, e.g.,][]{2023A&A...674A.179B}, and defines this of the density profiles. The 2D temperature profiles were deprojected and PSF-corrected following the non-parametric-like method from \cite{2010A&A...517A..52D} and \cite{2018A&A...617A..64B}. As shown in \cite{2024A&A...682A..45L}; De Luca et al. in prep., these temperature estimates are robust with respect to the employed pipeline. The temperature profile binning followed this adopted in \citet{2024A&A...686A..68R} for the CHEX-MATE DR1 sample, with the optimal binning scheme described in \citet{2024A&A...688A.219C}. 

Finally, assuming 
\begin{equation}
P_\mathrm{e}(r) = n_\mathrm{e}(r) \times k_\mathrm{B} T_\mathrm{e}(r), 
\end{equation}
the three dimensional pressure profile for each cluster was derived from the combination of the deprojected density, $n_\mathrm{e}(r)$, and the temperature, $T_\mathrm{e}(r)$, with $k_\mathrm{B}$ the Boltzmann constant. By construction, the pressure profiles are reconstructed in the same radial bins from the temperature and the resampled density profiles. The uncertainties on the pressure are propagated through Monte Carlo Gaussian realisation of the temperature and density profiles.
The pressure profiles extracted from \xmm\ data for the 28 clusters in the sample are shown in orange in Fig.~\ref{fig:pressureprofs}. To ensure a homogeneous radial coverage over our sample, we only consider \xmm\ profiles beyond $r/R_{500}=0.03$ (assuming the $R_{500}$ from MMF3 $M_{500}$ estimates, see Sect.~\ref{sec:m500s}). This way, we avoid biases in the core
that could be introduced by observational or physical effects \citep[signal-to-noise, presence of cool-cores, etc., see][and references therein]{2023AA...678A.197M}. The impact of the assumed centre on the pressure profile reconstruction is not investigated in this work, but it could introduce systematic shifts in the innermost bins for disturbed systems.

\subsubsection{\planck}
\label{sec:planckSZ}

Following the approach established for extracting the \planck\ tSZ signal in the PACT \citep[][]{2021AA...651A..73P} and X-COP \citep{2019AA...621A..41G,2019A&A...621A..39E,2022A&A...662A.123E} projects, in this work we used the non-public version of the MILCA \citep{2013A&A...558A.118H,2016A&A...594A..22P} all-sky $y$-map (Comptonisation parameter map), reconstructed at 7 arcmin full-width-half-maximum angular resolution.

The pressure profiles were recovered from the $y$-map following the method presented in \cite{2013AA...550A.131P} (hereafter P13) and used in subsequent works \citep{2019AA...621A..41G,2019A&A...621A..39E,2021AA...651A..73P,2022A&A...662A.123E}. In summary, Compton-$y$ profiles were computed by azimuthally averaging the signal in concentric annuli in the $y$-map centred at the X-ray peak of each cluster.
Point sources detected in the Planck Catalogue of Compact Sources \citep{2016A&A...594A..26P} and pixels with signal above $2.5\sigma$ noise level were masked (clipping).
The noise level was estimated from the dispersion of the signal in the area between $5$ to $10 \times \theta_{500}$  
around each cluster. A background zero-level of the $y$-map (defined from the average signal in $[5-10] \times \theta_{500}$) was removed from each individual Compton-$y$ profile. 

Compton-$y$ profiles were deprojected into three dimensional thermal pressure profiles by deconvolving from the \planck\ Gaussian PSF and applying the geometrical deprojection assuming spherical symmetry and adopting here again the method from \cite{2006A&A...459.1007C}. We propagated the 
contribution of the noise in the $y$-map and the correlation between the Compton-$y$ profile bins to the final pressure profile covariance matrices as done in previous works \citep{2013AA...550A.131P,2019AA...621A..41G,2019A&A...621A..39E,2021AA...651A..73P,2022A&A...662A.123E}. In short, $1000$ Compton-$y$ profile realisations computed from $1000$ noise realisation $y$-maps were used to estimate the noise covariance matrix associated to each data Compton-$y$ profile. Each noise covariance matrix was Cholesky decomposed to create $10000$ Gaussian realisations centred on each data Compton-$y$ profile. By performing the PSF deconvolution and deprojection of those $10000$ Compton-$y$ profiles, we obtained $10000$ pressure profiles used to compute each pressure covariance matrix ($\Sigma_{\mathrm{data}}$ in Eq.~\ref{eq:covmatrix}). Given the redshift range covered by the sample, some clusters are not fully resolved in the MILCA 7 arcmin $y$-map and their pressure profile bins are strongly correlated. The individual \planck\ pressure profiles for the DR1 CHEX-MATE clusters are shown in blue in Fig.~\ref{fig:pressureprofs}. Profiles are sampled on a regular grid with bins of width $\Delta r/R_{500} = 0.25$. We limit the profiles to $0.25 - 2.5 \times R_{500}$.

It should be noted that these \planck\ pressure profiles are not corrected for the relativistic contribution to the tSZ effect \citep{1998A&A...336...44P, 2006NCimB.121..487N}. Uncorrected tSZ signal is typically biased by $-5$ to $-15\%$ for $T_{\mathrm{e}} \sim 5-10$~keV galaxy clusters \citep[see][and De Luca et al. in prep.]{2019MNRAS.483.3459R, 2020MNRAS.493.3274L,2024PASA...41...87P}, depending on the temperature of each object. In Fig.~\ref{fig:pressureprofs}, we visually perceive a small offset between \xmm\ and \planck\ thermal pressure reconstructions within the common radial range. As quantified later in Sect.~\ref{sec:freeuppM500}, we find a typical ratio of $\sim - 5 \%$. Because the relativistic correction depends on cluster, and position within each cluster, it is non-trivial to include it within the MILCA maps. As a result, it is beyond the scope of this present work and tSZ data here are not corrected for relativistic contributions, just as in previous works \citep[e.g.,][]{2013AA...550A.131P,2019AA...621A..41G,2019A&A...621A..39E,2021AA...651A..73P,2022A&A...662A.123E}.

\subsection{$M_{500}$ estimates}
\label{sec:m500s}

As aforementioned, the masses used to scale the pressure are a key element in the reconstruction of the universal profile. In this paper, we will make use of different mass estimates. 

By `MMF3 masses' we refer to the estimates inferred from the tSZ signal measured by the multi-frequency matched filter MMF3 algorithm on \planck\ data \citep{2006A&A...459..341M,2014A&A...571A..29P}. These are the masses that were used for the selection of the CHEX-MATE sample \citep{2021A&A...650A.104C}. For each cluster, the amplitude of the tSZ signal $Y_{500}$ measured by the MMF3 algorithm was converted into an $M_{500}$ by applying the scaling relations calibrated beforehand on X-ray data \citep{2016A&A...594A..27P}.

In addition, 24 of the DR1 clusters have `dynamical mass' estimates. These have been derived from the velocity dispersion of their member galaxies in \citet{2025A&A...693A...2S}. These authors assumed as cluster centres the X-ray peaks from \cite{2023A&A...674A.179B} and adopted the modelings from \cite{1996ApJ...462..563N,2005MNRAS.363..705M,2019ApJ...871..168D}. Almost independent of the ICM properties, dynamical masses offer an estimate of the mass uncorrelated to the thermal pressure profiles used in this work. 
As for weak lensing masses, their systematic biases \citep{2005A&A...443..793G, 2021A&A...655A.115F} are generally contained down to a percent level \citep[e.g.,][]{2020A&A...641A..41F, 2021MNRAS.507.5671G, 2022A&A...659A.126A, 2024A&A...681A..67E, 2025A&A...697A.184G, 2025MNRAS.538L..50S}.
\cite{2025A&A...693A...2S} estimated a bias of $b_{\mathrm{PSZ2}} = M_{500}^{\mathrm{PSZ2}}/M_{500}^{\mathrm{dyn}} - 1 = - 0.38 \pm 0.04$ between dynamical and PSZ2 masses \citep[\planck\ 2nd Sunyaev-Zel'dovich catalogue,][]{2016A&A...594A..27P}. The agreement between the redshift estimates in  
\cite{2025A&A...693A...2S} to the redshifts considered in this work is very good: the difference of redshift estimates is $\Delta z / (1+z) < 0.004$ for every cluster. These 24 objects are very representative of the CHEX-MATE clusters studied by \cite{2025A&A...693A...2S} regarding the uncertainties of the individual dynamical masses (see Fig.~\ref{fig:dynerror}).

\section{Model}
\label{sec:model}

In this section, we detail the model that we use to describe the thermal pressure distribution in clusters. Following \cite{2007ApJ...668....1N} and \cite{2010AA...517A..92A}, we assume that for each cluster in the sample the pressure profile at the physical radius $r$ is defined as:
\begin{equation}
\label{eq:pressure}
    P(r) = P_{500} \times \mathbb{P}(x)
\end{equation}
with the normalised radius
\begin{equation}
\label{eq:r500}
    x \equiv r/R_{500}.
\end{equation}

Here $P_{500}$ is the normalisation factor of the pressure in physical units. We assume that it scales with mass and redshift following the self-similar model in Eq. A.1 in \cite{2010AA...517A..92A}, accounting also for a possible deviation from self-similar evolution with mass \citep[section 3.4 in][]{2010AA...517A..92A}:

\begin{eqnarray}
\label{eq:p500}
\nonumber P_{500}(M_{500}, z) = \frac{3}{8\pi} \left[ \frac{500 G^{-1/4}}{2} \right]^{4/3} \frac{\mu}{\mu_{\mathrm{e}}} f_{\mathrm{B}} \left[ M_{\mathrm{pivot}}  \right]^{ 2/3} \\
\times \; H(z)^{8/3}  \left[ \frac{M_{500}}{M_{\mathrm{pivot}}}  \right]^{\delta}.
\end{eqnarray} 

The $M_{500}$ mass is here in M$_{\odot}$ units and we take the pivot mass $M_{\mathrm{pivot}} = 3 \times 10^{14}\; \mathrm{M}_{\odot}$ to be consistent with \cite{2010AA...517A..92A}. We consider $\mu = 0.59, \mu_{\mathrm{e}}= 1.14,$ and $f_{\mathrm{B}} = 0.175$ for the mean molecular weight, the mean molecular weight per free electron, and the cosmic baryon fraction (defined as $f_{\mathrm{B}} = \Omega_{\mathrm{b}} / \Omega_{\mathrm{m}}$), respectively. These values are chosen to be identical to the ones used in \cite{2010AA...517A..92A} and \cite{2007ApJ...668....1N}. More recent estimates of the baryon fraction can be found in the literature \citep[$f_{\mathrm{B}} = \Omega_{\mathrm{b}} / \Omega_{\mathrm{m}} =  0.156 \pm 0.003$, $0.157 \pm 0.004$, and $0.157\pm 0.002$ from $\Omega_{\mathrm{b}}$ and $\Omega_{\mathrm{m}}$ in][respectively]{2016A&A...594A..13P,2020A&A...641A...6P,2024A&A...682A..37T}, the spread of these values and their uncertainties showing the current knowledge we have on $f_{\mathrm{B}}$. 
These newer constraints by \planck\ suggest a lower $f_{\mathrm{B}}$ than the value we adopted, which shifts the $P_{500}$ normalisation at the $5-10\%$ level. Regarding the molecular weights, we note that \cite{2019A&A...621A..40E} assumed $\mu= 0.61$ and $\mu_{\mathrm{e}} = 1.13$, while \cite{2009A&A...501...61E} considered $\mu= 0.600$ and $\mu_{\mathrm{e}} = 1.155$. Molecular weight values depend on the adopted metallicity, and so on the assumed relative abundance table.

In Eq.~\ref{eq:p500}, the scaling of $P_{500}$ with $M_{500}$ is determined by the power $\delta$, which in the gravitation-only self-similar scenario takes $\delta = 2/3$. In our framework, $\delta$ can also be a free parameter of the model. We assume the redshift evolution given by the gravitation-only self-similar model, $P_{500}\propto H(z)^{8/3}$. As already investigated in \citet{2023ApJ...944..221S,2012ApJ...758...75B}, and \citet{2014ApJ...794...67M} for pressure profiles  \citep[and in][for gas density profiles]{2022A&A...665A..24P}, there could be a deviation from the self-similar evolution with redshift. Since we do not account for it, if any, this will have an impact on our measurement of the evolution of $P_{500}$ with mass.

The second term in Eq.~\ref{eq:pressure} corresponds to the dimensionless UPP that we aim at fitting. It is meant to describe, on average, the shape of the scaled radial pressure profiles in galaxy clusters. We parametrise $\mathbb{P}(x)$ following the gNFW model:
\begin{equation}
\label{eq:gnfw}
    \mathbb{P}(x) = \frac{P_0}{(c_{500}x)^{\gamma} \left[1+(c_{500}x)^{\alpha}\right]^{(\beta-\gamma)/\alpha}},
\end{equation}
where $P_0$ is the normalisation constant, $c_{500}$ the concentration, $\beta$ and $\gamma$ are the outer and central power law exponents, respectively, and 
$\alpha$ gives the slope transition steepness.

Previous works in the literature \citep[e.g.,][]{2019AA...621A..41G,2023ApJ...944..221S} have shown the need to account for an intrinsic scatter associated to the universal profile. This scatter quantifies the spread about the UPP across the cluster population due to the variety of dynamical states as well as to the impact of non-gravitational processes governing the baryon physics. We assume a log-normal intrinsic scatter profile $\sigma_{\mathrm{int}}(x)$ \citep[see Eq.~\ref{eq:scattermatrix} and][]{2019AA...621A..41G,2023A&A...674A.179B}
and choose to parametrise it as a function of the scaled radius $x$, following:
\begin{equation}
\label{eq:ourscatter}
    \sigma_{\mathrm{int}} (x) =  \sigma_1 \exp{\left[- \omega x\right]} + \sigma_0 x,
\end{equation}
with $\sigma_1, \sigma_0$, and $\omega$ the free parameters in the model. In Appendix~\ref{sec:intrinscatter} we discuss the choice of this particular parametrisation with respect to other options in the literature. We can also account for the covariance of the Gaussian process describing the intrinsic scatter with an additional free parameter $L_{\mathrm{int}}$ as detailed in Sect.~\ref{sec:fittingmethod} (Eq.~\ref{eq:intcovmat}).

In addition, when combining X-ray and tSZ-based pressure profiles one must account for a potential systematic discrepancy between them \citep[][De Luca et al. in prep.]{2019A&A...621A..34K}. As investigated in \cite{2019A&A...621A..34K,2020AN....341..210E,2021MNRAS.504.1062W}, multiple factors can be at the origin of such a discrepancy and, in the literature, it has been quantified as $\mathcal{R}$ or $\eta_T$:
\begin{equation}
    \eta_T \sim P^{\mathrm{X}}/P^{\mathrm{tSZ}}.
\end{equation}
We consider the normalisation factor $\eta_T$ a free parameter that calibrates the ratio between X-ray and tSZ-based pressure profiles. 

In summary, our model contains two types of free parameters, $\theta$:
\begin{itemize}
    \item Global parameters, that is, parameters that are common to all clusters. These are: the power in the $P_{500}-M_{500}$ scaling relation ($\delta$); the parameters of the UPP gNFW profile $(P_0, c_{500}, \alpha, \beta, \gamma )$; the parameters describing the intrinsic scatter ($\sigma_1, \sigma_0, \omega$, and $L_{\mathrm{int}}$); and the ratio between X-ray and tSZ pressure profiles ($\eta_T$). 
    \item Individual parameters. In particular, we have one mass parameter per galaxy cluster: $\{ M_{500, 1}, ..., M_{500, n} \}$. In a fixed cosmology, and assuming a redshift per cluster, the conversion between $M_{500}$ and $R_{500}$ (needed to calculate the $x$ in the model, Eq.~\ref{eq:r500}) is $\mathrm{straightforward}^{\ref{M500def}}$.
\end{itemize}
In this framework, any of the mentioned parameters can be fitted or fixed to a predefined value.

In the literature, the power law exponent in the $P_{500}-M_{500}$ scaling relation is often fixed to the self-similar $\delta_0=2/3$ value \citep[e.g.,][]{2019AA...621A..41G}. Nevertheless, some works \citep{2010AA...517A..92A, 2013ApJ...768..177S,2021ApJ...908...91H, 2023ApJ...944..221S} have already studied the deviation from self-similarity, obtaining values that span from
$\delta \sim \delta_0-0.20$ to $\delta \sim \delta_0+0.12$. According to simulations \citep{2017MNRAS.466.4442L, 2018MNRAS.480.2898C}, the power should be very close to $\delta_0=2/3$, but reaching almost $\delta_0+0.33$ in \cite{2017MNRAS.466.4442L} and $\delta_0-0.03$ in \cite{2018MNRAS.480.2898C}. In addition, as shown in \cite{2021ApJ...908...91H}, the estimation of $\delta$ is correlated with the assumed mass scale.

Regarding the intrinsic scatter of the universal profile, we find in the literature works that fit jointly the scatter and the pressure profile \citep{2013ApJ...768..177S,2019AA...621A..41G, 2023ApJ...944..221S} and others that measure a posteriori the spread of the data with respect to the UPP \citep{2010AA...517A..92A,2017ApJ...843...72B}. We will here opt for the joint fit.

For the very first time with respect to other works, in this paper we allow individual cluster masses to vary along the fit of the universal pressure profile. This constitutes the main originality of the presented approach.

\section{Fitting method}
\label{sec:fittingmethod}

In practice, our model assumes that each individual pressure profile $P_{\mathrm{data}, i}$ (Fig.~\ref{fig:pressureprofs}) is well described by a multivariate Gaussian distribution 
\begin{equation}
    \mathcal{L}_i = \frac{1}{\sqrt{(2 \pi)^{n_i} |C_i|}} \mathrm{exp}\left[ -\frac{1}{2} (P_{\mathrm{data}, i}- P_{\mathrm{mod}, i})^T C_i^{-1} (P_{\mathrm{data}, i}- P_{\mathrm{mod}, i}) \right]
\end{equation}
with $n_i$ the number of data pressure bins for cluster $i$. In the above equation, $P_{\mathrm{mod}, i}(\theta)$ corresponds to the pressure profile model (Eq.~\ref{eq:pressure}) and $C_i(\theta)$ is the covariance matrix \citep{2013ApJ...768..177S, 2023ApJ...944..221S,2023A&A...674A.179B} given by: 
\begin{equation}
    \label{eq:covmatrix}
    C_i(\theta) = \Sigma_{\mathrm{data}, i} + \Sigma_{\mathrm{int},i}(\theta),
\end{equation}
where $\Sigma_{\mathrm{data}, i}^{k, l}$ is the covariance of the data pressure profile for the radial bins $k$ and $l$. In this equation, $\Sigma_{\mathrm{int}, i}$ is a matrix that accounts for the intrinsic dispersion. 
Since we consider a log-normal scatter, each diagonal element of $\Sigma_{\mathrm{int}, i}$ is defined as
\begin{equation}
    \label{eq:scattermatrix}
    \begin{split}
    \Sigma_{\mathrm{int}, i}^{k, k} (\theta) & =  
    \left[ P_{\mathrm{mod}, i} (r_{i, k},  \theta) \times  \sigma_{\mathrm{int}} (r_{i, k}, \theta) \right]^2  \\
    & = \left[ P_{500, i} (\theta) \;  \mathbb{P}(r_{i, k},  \theta) \times  \sigma_{\mathrm{int}} (r_{i, k}, \theta) \right]^2.
    \end{split}
\end{equation}

Here $P_{500, i}$ (Eq.~\ref{eq:p500}), $\mathbb{P}$ (Eq.~\ref{eq:gnfw}) and $\sigma_{\mathrm{int}}$ (Eq.~\ref{eq:ourscatter}) are functions of the free parameters in the model: $\theta = \{ M_{500, 1}, ..., M_{500, n}; P_0, c_{500}, \alpha, \beta, \gamma; \delta; \eta_T; $, $\sigma_1, \sigma_0, \omega$\}. In Eq.~ \ref{eq:scattermatrix}, $r_{i, k}$ corresponds to the physical radius of the pressure profile for the $k$-th bin in cluster $i$. 

Pressure profiles are thought to be
largely smooth, neighbouring points correlating with each other. Thus, we introduce a correlation in the intrinsic scatter covariance matrix considering a ``radial basis function'' or ``squared-exponential'' kernel
\citep[as already used, e.g., in][]{2021MNRAS.501.1463K}, the kernel size being defined by the $L_{\mathrm{int}}$ parameter: 
\begin{equation}
    \Sigma_{\mathrm{int}, i}^{k, l} (\theta) = \sqrt{\Sigma_{\mathrm{int}, i}^{k, k} \Sigma_{\mathrm{int}, i}^{l, l}} \exp{\left[  - \frac{(x_{i, k} - x_{i, l})^2}{2 L_{\mathrm{int}}^2} \right]}, 
    \label{eq:intcovmat}
\end{equation}
with $x_{i, k} = r_{i, k}/R_{500, i}$ and $x_{i, l} = r_{i, l}/R_{500, i}$ following Eq.~\ref{eq:r500}. By considering this covariance kernel for the intrinsic scatter, we assume contiguous pressure bins to be more correlated to each other and no anticorrelations. The $L_{\mathrm{int}}$ parameter quantifies the radial distance between correlated bins, and it allows modelling a systematic deviation with respect to the UPP.

Markov chain Monte Carlo (MCMC) sampling is used to fit the free parameters describing the model to the data. Given the large number of parameters that the model can encompass,
an efficient fitting method is required. For this reason, we rely on \texttt{jax}-based codes \citep{jax2018github} that enable the automatic differentiation of distributions, and thus, a fast sampling of the posterior probability distribution. We perform the Bayesian inference making use of the no U-turn sampler \citep[NUTS,][]{JMLR:v15:hoffman14a} as implemented in the \texttt{numpyro} \citep{bingham2019pyro, 2019arXiv191211554P, 2025ascl.soft05005P} \texttt{python} library. In the Bayesian framework, our posterior probability distribution of the parameters is a combination of the multivariate Gaussian distribution quantifying the likelihood of the data to satisfy the model and the prior distribution associated to each free parameter. The final likelihood is the product of all the $n$
individual cluster likelihoods: $\mathcal{L} = \prod_{i=1}^{n} \mathcal{L}_i $. Table~\ref{tab:parampriors} summarises the priors considered in this work. 

\renewcommand{\arraystretch}{1.4}   
\begin{table}[]
    \caption{Prior distributions for the parameters in the model.
    } 
    \centering
    \begin{tabular}{c c}
    \hline\hline
        Parameters  &  Prior distribution \\  \hline
        
        $P_0$   &  $\mathcal{U}(0, 60)$ \\ 
        $c_{500}$ &   $\mathcal{U}(0, 10)$ \\
        $\alpha$ &  $\mathcal{U}(0, 10)$  \\
        $\beta$ &  $\mathcal{U}(0, 20)$\\ 
        $\gamma$ &   $\mathcal{U}(0, 10)$ \\
        $\delta$  &   $\mathcal{N}(2/3, 0.2^2)$ \\
        $\eta_T$     & $\mathcal{U}(0, 4)$ \\
        $L_{\mathrm{int}}$     & $\mathcal{U}(0, 2.5)$ \\
        $\sigma_{0}$ &  $\mathcal{U}(0, 3)$   \\
        $\sigma_{1}$ &  $\mathcal{U}(0, 10)$ \\
        $\omega$ &    $\mathcal{U}(-2, 10)$ \\
        $M_{500, i}$ & $\mathcal{N} \left( M_{500, i}, \sigma^2_{ M_{500, i}}  \right) H(0)$\\ \hline   
    \end{tabular}
    \tablefoot{For each cluster $i$ we assume a normal distribution on the mass based on $M_{500, i}$ masses from catalogues (Sect.~\ref{sec:m500s}). All masses are truncated at $M_{500, i} =0$ as indicated by the Heaviside function $H(0)$. 
    } 
    \label{tab:parampriors}
\end{table}

The sampling is performed using 10 chains, $10^3$ tuning steps and the sampling steps needed to reach convergence (typically $\sim 10^5$ or $\sim 10^4$ depending if individual cluster masses are fitted or not). We keep the four chains that have reached the highest posterior probability distributions and consider that MCMC chains have converged if the $\hat{R} < 1.01$ condition \citep{1992StaSc...7..457G,2021BayAn..16..667V} is reached.

\section{Measuring the universal pressure profile and individual cluster masses} 
\label{sec:data}

\begin{figure*}[h]
    \centering
    \includegraphics[scale=0.4]{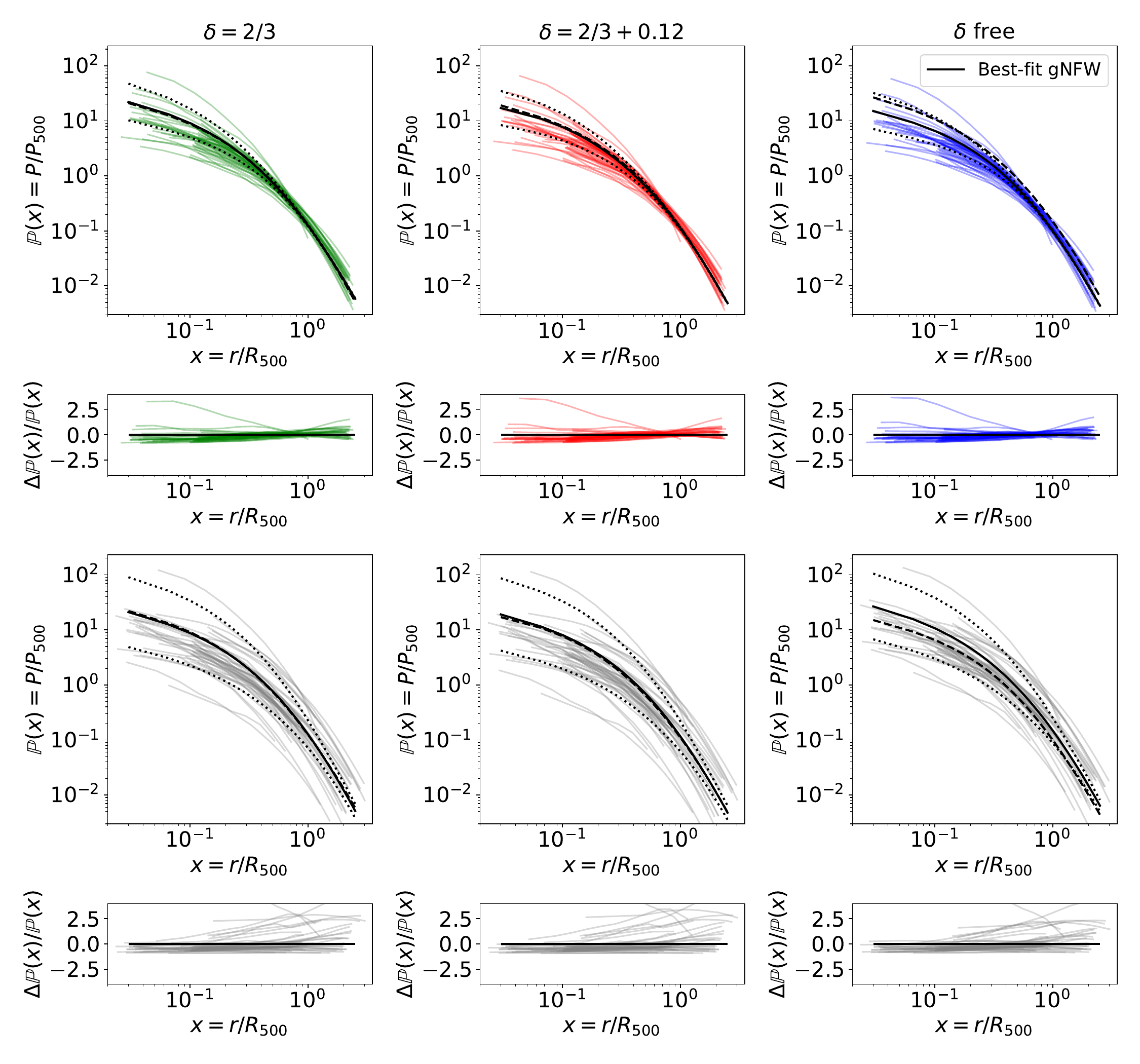}
    \caption{Normalised data pressure profiles and best-fit gNFW models, and their relative difference. Top: coloured profiles show the individual data pressure profiles (from Fig.~\ref{fig:pressureprofs}) for the 24 clusters normalised by the best $P_{500, i}$ and $R_{500, i}$ when fitting the individual masses with dynamical mass estimates as priors. Green, red, and blue correspond to the fits assuming respectively $\delta=2/3$, $\delta=2/3+0.12$, and $\delta$ as a free parameter. Bottom: in grey data pressure profiles normalised by $P_{500, i}$ and $R_{500, i}$ considering the masses fixed in the fit and setting them to the dynamical mass estimates. The solid black profile in each panel indicates the best-fit gNFW profile for that case, with the dotted profiles showing the fitted intrinsic scatter $ \mathbb{P}(x) \exp[\pm \sigma_{\mathrm{int}}(x)]$. The dashed profiles in the top (bottom) panels correspond to the best-fit gNFW for the mass fixed (fitted) cases.
    }
    \label{fig:dynamicalfit}
\end{figure*}

In this section, we present the results obtained by applying the described method to our CHEX-MATE data set (Sect.~\ref{sec:dataset}) for the 24 clusters with available dynamical masses.

We consider all the individual cluster $M_{500, i}$ masses as free parameters. This means that, at each step of the fit, both $P_{500}$ and $x$ values will vary, without adapting the number of considered pressure bins at each step. The change is accounted for in the computation of the model at each step of the MCMC and not on the data. A shift in $M_{500}$ and, thus, in $R_{500}$, implies therefore a normalisation change of the $P(r)$ pressure profile in the ordinate (Eq.~\ref{eq:p500}) as well as in the abscissa ($r$ in Eq.~\ref{eq:r500}), shrinking or stretching the profile in both axes. Such a shift in $M_{500}$ can mimic the effect of a modification in gNFW shape parameters. This means that when pressure profiles are fitted by assuming fixed $M_{500}$ values, the resulting UPP is directly affected by the considered masses. And as a consequence, the use of inadequate mass estimates will lead to wrong UPP measurements.

Dynamical mass estimates (Sect.~\ref{sec:m500s}) are ideal priors for the cluster masses in the joint fit of our data pressure profiles: they are expected to be scattered but (almost) unbiased mass estimates \citep{2005A&A...443..793G,2021A&A...655A.115F}. Appendix~\ref{sec:simulations} presents the validation of our method and shows that, if unbiased, even scattered mass estimates are good priors to inform the fit. Moreover, dynamical masses are mostly uncorrelated to the \xmm\ and \planck\ pressure profiles (Sect.~\ref{sec:pressprof}).

\subsection{Joint fit of individual cluster masses and the UPP}
\label{sec:freeuppM500}

By taking Gaussian priors centred on the dynamical masses with their deviation corresponding to the dynamical mass measurement errors (Table~\ref{tab:parampriors}) as estimated in \cite{2025A&A...693A...2S}, we perform the joint fit of the universal gNFW pressure profile and the individual masses for the 24 clusters of our subsample (see Sect.~\ref{sec:m500s}).

\renewcommand{\arraystretch}{1.4}   
\begin{table*}[]
    \caption{Best-fit parameters from the joint fit of pressure profiles assuming dynamical mass priors and for different $\delta$ values.
    }
    \centering
    \begin{tabular}{c c c c }
    \hline\hline
        Parameter  &   $\delta=2/3$ &  $\delta=2/3+0.12$ &  $\delta$ free \\  \hline
        
        $P_0$   & $16.63 \pm 10.38 \; (11.40)$  & $14.21 \pm 8.85 \; (25.63)$ &  $16.19 \pm 10.17 \; (5.55)$ \\

        $c_{500}$ & $1.73 \pm 0.58 \; (1.69)$ & $1.76 \pm 0.58 \; (1.21)$ &  $1.77 \pm 0.56 \; (1.68)$ \\

        $\alpha$ & $0.95 \pm 0.20 \; (0.94)$ &  $0.96 \pm 0.20 \; (0.76)$ &  $0.96 \pm 0.20 \; (1.07)$\\

        $\beta$ &  $4.69 \pm 0.94 \; (4.54)$ & $4.65 \pm 0.87 \; (5.46)$ & $4.62 \pm 0.82 \; (4.47)$  \\

        $\gamma$ & $0.24 \pm 0.14 \; (0.31)$  & $0.24 \pm 0.14 \; (0.04)$ & $0.24 \pm 0.14 \; (0.38)$ \\

        $\delta$  & - & - & $0.65 \pm 0.12 \; (0.85)$  \\

        $\eta_T$ & $1.06 \pm 0.02 \; (1.05)$ & $1.06 \pm 0.02 \; (1.05)$ &  $1.06 \pm 0.02 \; (1.04)$\\

        $L_{\mathrm{int}}$ & $0.30 \pm 0.02\; (0.28)$ &  $0.30 \pm 0.02 \; (0.27)$ & $0.30 \pm 0.02 \; (0.27)$\\

        $\sigma_{0} \; [\times 10^{-2}]$ & $1.05 \pm 0.32 \; (0.98)$ & $1.06 \pm 0.33 \; (0.80)$  & $1.05 \pm 0.32 \; (0.89)$  \\

        $\sigma_{1}$ & $0.88 \pm 0.14 \; (0.85)$  & $0.87 \pm 0.14 \; (0.79)$ & $0.88 \pm 0.14 \; (0.85)$ \\

        $\omega$ &  $3.04 \pm 0.35 \; (3.49)$ &  $3.04 \pm 0.36 \; (3.37)$ &  $3.03 \pm 0.35 \; (3.79)$ \\\hline
       
    \end{tabular}
    \tablefoot{For each parameter, we give: the mean value and its standard deviation calculated from the marginalised posterior distribution, and in parenthesis, the best-fit value.
    }
    \label{tab:fitted params}
\end{table*}

\begin{figure}
    \centering
    \includegraphics[scale=0.45]{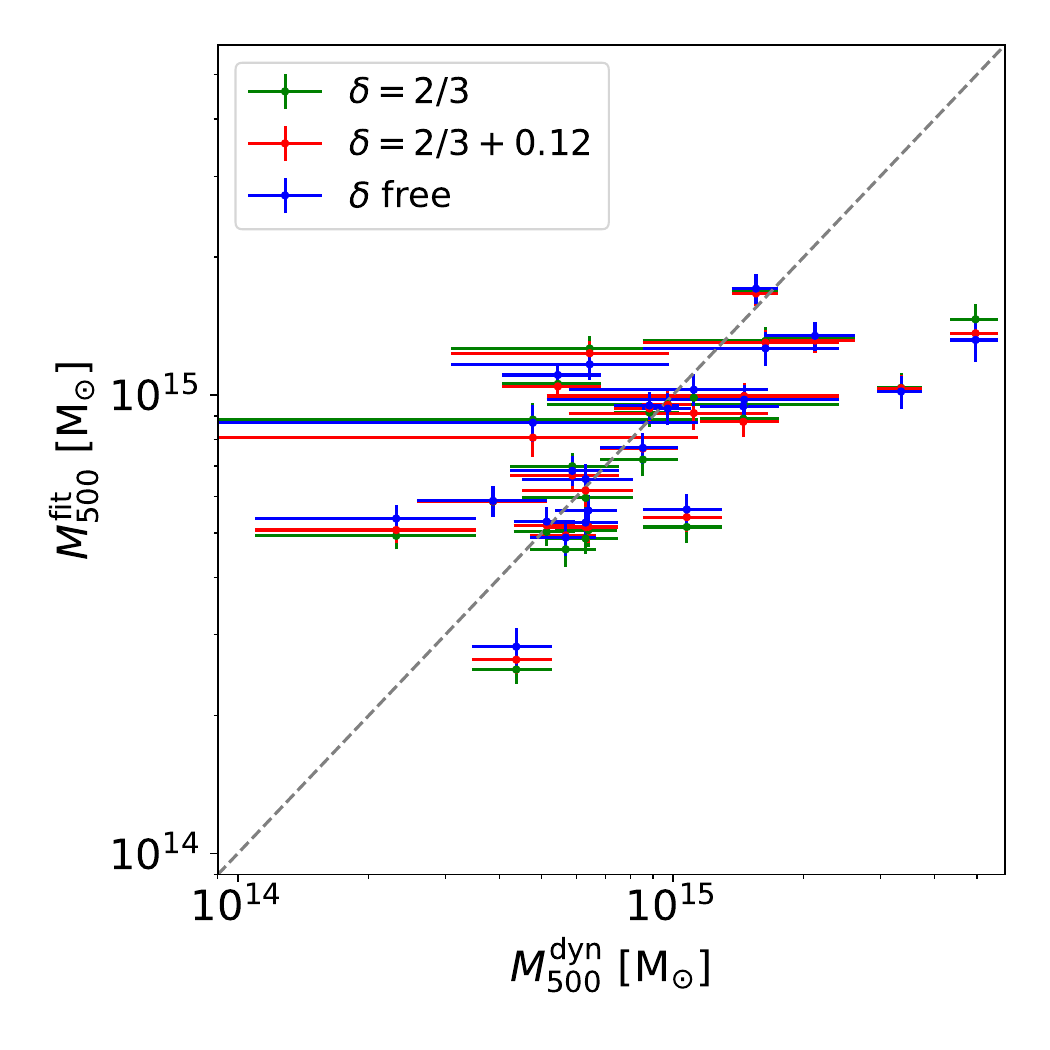}
    \includegraphics[scale=0.45]{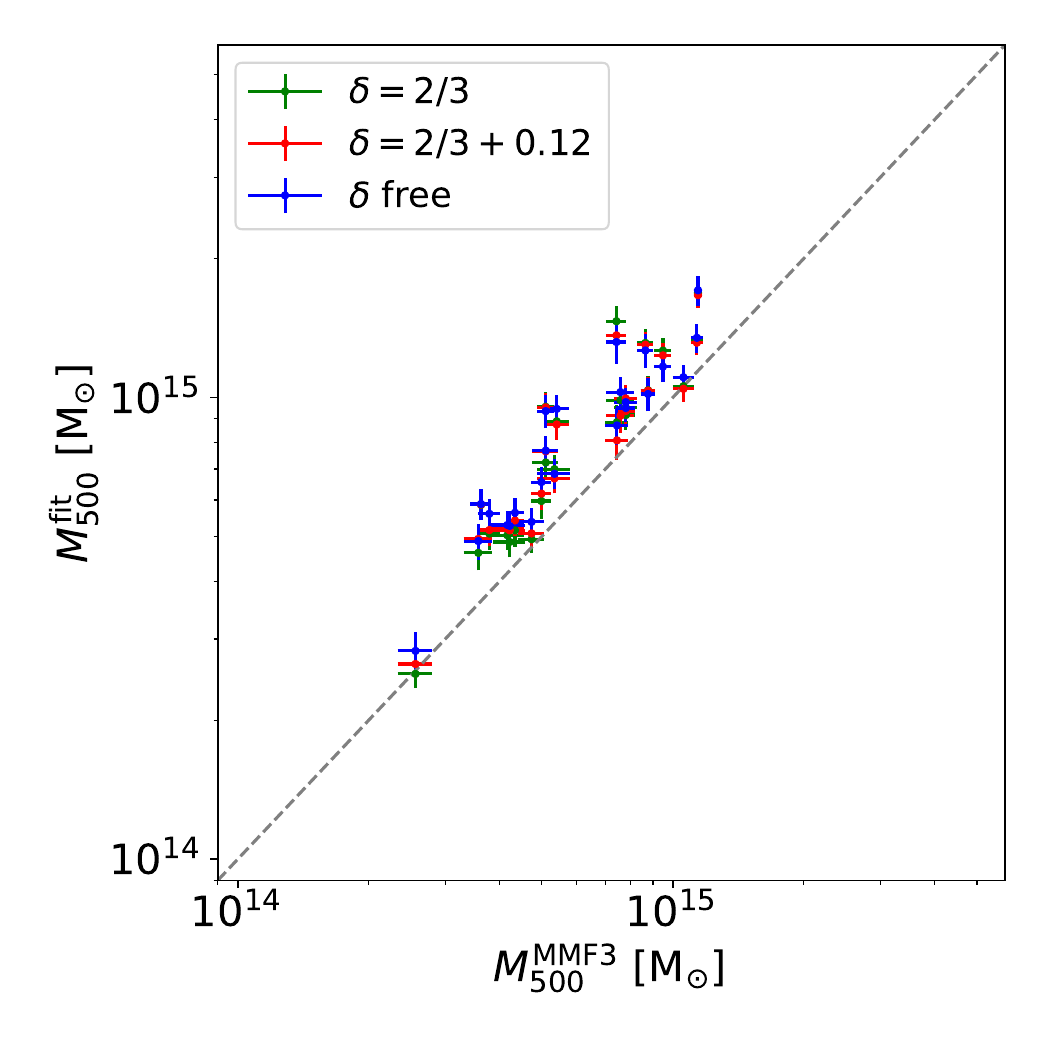}
    \caption{Masses obtained from the joint fit to data by taking Gaussian priors centred on dynamical estimates. We compare the results to dynamical (top) and MMF3 (bottom) mass estimates. The colour scheme is identical as in Fig.~\ref{fig:dynamicalfit} for the different values of $\delta$. Uncertainties of fitted masses are calculated as the standard deviation of the marginalised posterior distribution for each cluster mass parameter.
    }
    \label{fig:dynamicalfitM500}
\end{figure}

\begin{figure}[h!]
    \centering
    \includegraphics[scale=0.44, trim={0cm 1cm 0cm 0cm}]{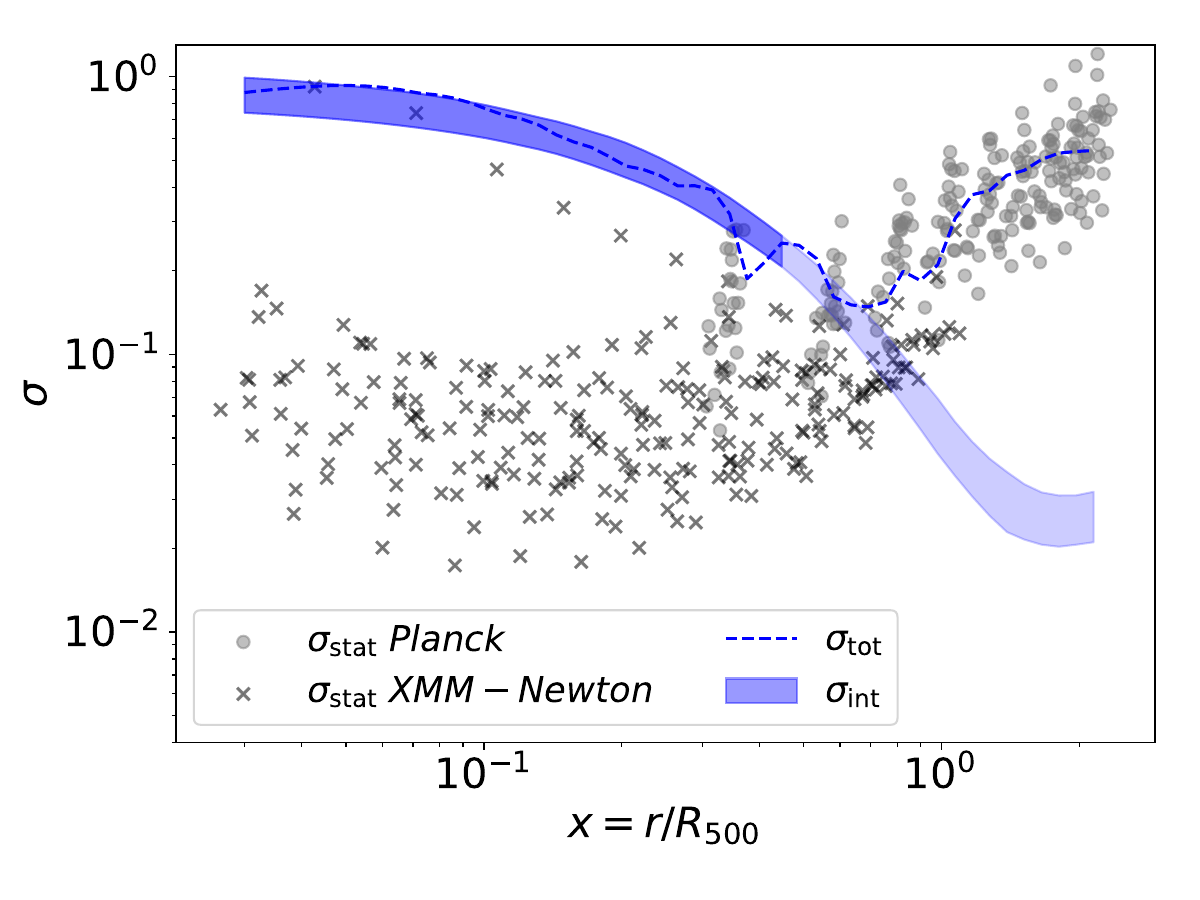}
    \caption{Scatter of pressure profiles for the 24 clusters in our sample. Shaded area indicates 16th to 84th percentiles for the intrinsic scatter profile fitted to data pressure profiles ($\delta$ free case). The intrinsic scatter may not be reliably constrained beyond $x \sim 0.4$. The dashed line shows the scatter of the individual normalised profiles with respect to the best-fit gNFW model. Statistical uncertainties of the individual pressure profiles are indicated with crosses and circles for the pressure bins corresponding to \xmm\ and \planck\ data, respectively. All scatters are given in $P_{500} \mathbb{P}(x)$ units.
    }
    \label{fig:scatter_fits}
\end{figure}

We consider three cases: a first one in which $\delta$ is fixed to the self-similar value, a second in which it is fixed to $\delta=2/3+0.12$ \citep[as in][]{2010AA...517A..92A}, and a third case where $\delta$ is a free parameter of the model. For the latter, we consider a Gaussian prior distribution centred on the self-similar $2/3$ with a standard deviation of 0.2 (Table~\ref{tab:parampriors}). By investigating these three cases, we study the impact of the assumed $P_{500}-M_{500}$ scaling relation on the UPP.
We adopt flat priors for the rest of the parameters (Table~\ref{tab:parampriors}).

Colour profiles in top panels in Fig.~\ref{fig:dynamicalfit} show the normalised individual pressure profiles for our 24 galaxy clusters. They are normalised by the $P_{500, i}$ and $R_{500, i}$ for the best-fitting $M_{500, i}$ and $\delta$ parameters (Table~\ref{tab:fitted params}). Solid black profiles show the best-fitting gNFW model for each case of $\delta$ (parameter values are summarised in Table~\ref{tab:fitted params}), and below each panel we present the relative difference between the normalised individual profiles and the best-fitting gNFW. The corner plot in Fig.~\ref{fig:cornerall} shows the 1D and 2D posterior distributions for all the fitted parameters, colour coded as in Fig.~\ref{fig:dynamicalfit}. Regarding the ratio between X-ray and tSZ data, we obtain for all three cases a best-fit value of $\eta_T \sim 1.05$, consistent with what would be expected from neglecting tSZ relativistic corrections (Sect.~\ref{sec:planckSZ}).

For what concerns the power law in the $P_{500}-M_{500}$ scaling relation, a high value of $\delta = 0.85$ 
is preferred by the data when we let this parameter free, in line with \cite{2010AA...517A..92A,2020A&A...644A.111E,2023A&A...669A.133E}. However, the marginalised posterior distribution of $\delta$ is in agreement with self-similar evolution. As expected, $\delta$ is mainly correlated to $P_0$ and cluster masses (Fig.~\ref{fig:cornerall}). The departure of the best-fit $\delta$ from the self-similar expectation could be interpreted as a mass dependent gas mass fraction $f_{\mathrm{gas}} \propto M_{500}^{\delta -2/3}$ \citep{2020A&A...644A.111E,2023A&A...669A.133E}, our results suggesting an increase of the gas mass fraction with cluster mass \citep[see other recent results in][]{2023A&A...674A..48W,2023A&A...669A.133E}. This trend could be
linked to non-gravitational processes that can modify the ICM thermodynamics, such as AGN feedback and turbulence \citep{2020MNRAS.498.4983W}. The model presented in this work could be extended to directly fit $f_{\mathrm{gas}}$ in future applications. It is also worth noting the small impact of the various $\delta$ values considered in this work on the rest of the free parameters.

The best-fitting individual cluster masses are presented with respect to the dynamical estimates in the top panel in Fig.~\ref{fig:dynamicalfitM500}. Green, red, and blue markers correspond to the masses obtained by assuming different power values in the $P_{500}-M_{500}$ scaling relation. We observe that, by construction of the priors, fitted masses are scaled to the dynamical mass scale ($<M_{500}^{\mathrm{fit}}/M_{500}^{\mathrm{dyn}} > = 1.004 \pm 0.507$, $1.003\pm 0.494$, and $1.031 \pm 0.513$ for $\delta=2/3, 2/3+0.12$, and $\delta$ free cases, respectively), but corrected when informed by the thermal pressure distribution. 
We stress that in Fig.~\ref{fig:dynamicalfitM500} the error bars of the fitted masses correspond to the standard deviation of the marginalised posterior distribution for each cluster mass parameter. These error bars are not representative of the uncertainties we have on individual masses since they are all strongly correlated to each other. The corner plot in Fig.~\ref{fig:cornerall} illustrates these correlations between the $M_{500, i}$. Fig.~\ref{fig:cornermass} compares the marginalised posterior distributions of the 24 $M_{500, i}$ parameters to the Gaussian priors assumed for the cluster masses, showing how fitted masses deviate from the considered prior distributions. In the bottom panel in Fig.~\ref{fig:dynamicalfitM500}, we compare the fitted masses to the MMF3 estimates. They show a very good correlation. On average, their ratio is of $<M_{500}^{\mathrm{fit}}/M_{500}^{\mathrm{MMF3}} > = 1.325 \pm 0.245$, $1.327 \pm 0.228$, and $1.356 \pm 0.212$ (for $\delta=2/3, 2/3+0.12$, and $\delta$ free cases, respectively), compatible with the bias obtained in \citet{2025A&A...693A...2S}. Similarly, when comparing fitted masses to HSE estimates reconstructed from \xmm\ data, we obtain $< M_{500}^{\mathrm{fit}}/M_{500}^{\mathrm{XMM}} > = 1.416 \pm 0.332$, $1.419 \pm 0.321$, and $1.447 \pm 0.293$ respectively for $\delta=2/3, 2/3+0.12$, and $\delta$ free cases.

By jointly fitting the universal pressure profile and cluster masses, we have propagated the uncertainties on the individual masses to the UPP. In addition, 
we recover the intrinsic scatter of the pressure profiles, due mainly to the baryon physics and variations in clusters dynamical states across our sample, without being impacted by the scatter of the mass estimates used to scale the individual profiles \citep[see][]{2019AA...621A..41G,2024A&A...691A.340R}. The shaded profile 
in Fig.~\ref{fig:scatter_fits} shows the fitted intrinsic scatter profile.
We only present the profile obtained from the fit with $\delta$ as a free parameter, but $\delta=2/3$ and $\delta=2/3+0.12$ cases give fully compatible results (see Fig.~\ref{fig:cornerall}).
We observe a decreasing trend with radius up to $x \sim 1.2$, where the scatter starts to rise again. In the same figure, the dashed profile indicates the scatter of the individual pressure profiles with respect to the UPP ($\sigma_{\mathrm{tot}}$). Grey crosses and circles show the statistical uncertainties ($\sigma_{\mathrm{stat}}$, square root of the covariance matrix diagonal) of each individual pressure profile for the \xmm\ and \planck\ bins, respectively. We notice that at large radii (beyond $x\sim 0.4$) the level of statistical uncertainties prevents us from constraining the underlying intrinsic scatter\footnote{The tight constraints we obtain on $\sigma_0$, $\sigma_1$, and $\omega$ are an artefact of the chosen parametrisation, which does not allow for a low intrinsic scatter profile and large uncertainties.}: statistical uncertainties are as large as $\sigma_{\mathrm{tot}}$, which explains the very low intrinsic scatter we measure compared to other works (Fig.~\ref{fig:intrinscatter}). We have verified that the same effect is observed if we perform the joint fits by modelling the intrinsic scatter with the parametrisation in Eq.~\ref{eq:ghirardini} from \cite{2019AA...621A..41G}, and that fitted $M_{500, i}$ parameters 
remain consistent. 
By letting $L_{\mathrm{int}}$ free in the fit, we measure a correlation length of $\sim 0.3$.

We repeated the fits but fixing the cluster masses to the dynamical estimates in \cite{2025A&A...693A...2S}. The individual normalised pressure profiles and best-fit gNFW profiles are shown in the bottom panels in Fig.~\ref{fig:dynamicalfit}. Unsurprisingly, given their large scatter, dynamical masses are not suited to scale directly the individual profiles, leading to an overestimation of the intrinsic scatter. On the contrary, by jointly fitting the pressure and masses, we are able to refine the latter and find the masses that scale the pressure profiles the best (top panels in Fig.~\ref{fig:dynamicalfit}). 

Finally, we performed the joint fit of individual masses and UPP parameters
by considering only the \xmm\ pressure bins (orange profiles in Fig.~\ref{fig:pressureprofs}). Given the radially constant and
small statistical uncertainty sizes (Fig.~\ref{fig:scatter_fits}), we recover the intrinsic scatter profile shape expected from previous works (see Fig.~\ref{fig:intrinscatter}).
However, the radial extent of \xmm\ profiles is not large enough to fully constrain the gNFW profile shape and significantly improve on the mass estimates. Consequently, the intrinsic scatter is overestimated by the dispersion of dynamical mass estimates.

Further analyses could try to model the intrinsic scatter in the $P_{500}-M_{500}$ scaling relation, as well as the scatter of the fitted masses with respect to the true $M_{500}$, specially if mass proxies with small nominal uncertainties are used as priors. The current work presumes those scatters to be zero. Although the impact of the assumed cosmology on the resulting UPP and masses is not assessed in this paper either, it directly affects the results through the deprojection of the individual pressure profiles and definitions of $R_{500}$ and $P_{500}$ (Eq.~\ref{eq:p500}). This impact should be quantified and coherently considered for further use in cosmological analyses.

\subsection{
Impact of the UPP on the mass determination}
\label{sec:fixeduppfits}

\begin{figure}
    \centering
    \includegraphics[scale=0.41, trim={1.5cm 0cm 0cm 0cm}]{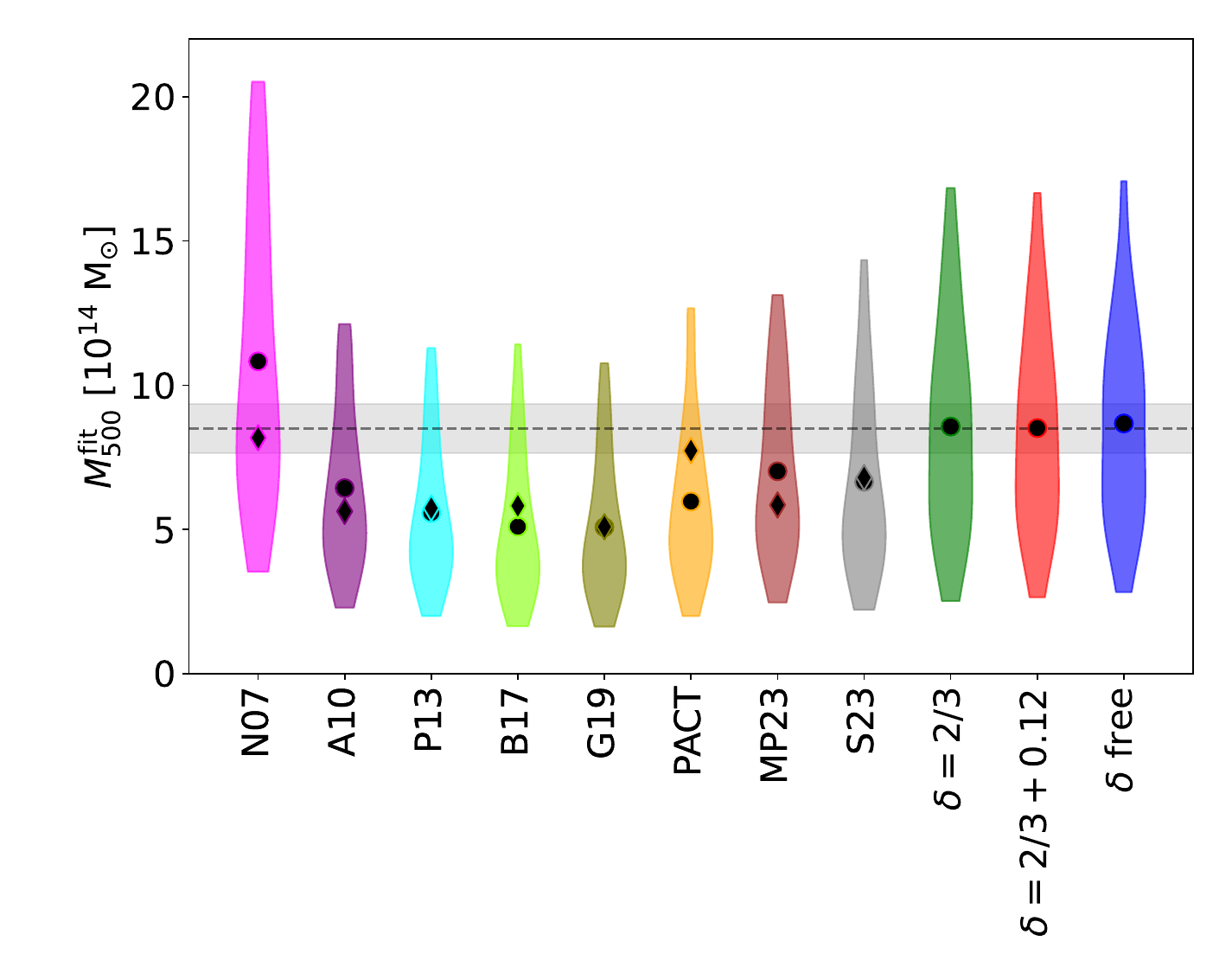}
    \caption{Distribution of the 24 best-fit $M_{500}$ obtained from the thermal pressure profile fits when $\eta_T$ is let free in the fit. We present the average mass along the sample for fixed UPPs from the literature with circles and diamonds for the cases where $\eta_T$ is let free in the fit and fixed to $\eta_T = 1.05$, respectively. We also give the $M_{500}^{\mathrm{fit}}$ distributions and average masses of the sample obtained from the joint fits of the UPP and individual cluster masses, with $\eta_T$ and $\sigma_{\mathrm{int}}$ free in the fits. Different colours correspond to results obtained with the parameters specified in Table~\ref{tab:uppliterature} and our three results assuming different $\delta$ values.
    The horizontal dashed line indicates $8.5 \times 10^{14}$ M$_{\odot}$ as a reference, with the shaded area corresponding to a $10\%$ dispersion around this value. 
    }
    \label{fig:averageM500}
\end{figure}

In the previous sections (and in Appendix~\ref{sec:simulations}), we have shown the complex degeneracies between the parameters in the model.
Here, we further stress the intricate inter-dependencies between the UPP and cluster masses, by assessing the impact of the UPP shape on the measured masses over a given sample.

We considered various gNFW profiles from the literature
\citep[N07, A10, P13, B17, G19, PACT, S23, and][hereafter MP23]{2023AA...678A.197M}, assuming also the associated values of $\delta$ adopted in each of these studies (Table~\ref{tab:uppliterature}). These eight gNFW profiles are drawn in Fig.~\ref{fig:uppliterature}. For each case, we fixed the given universal pressure profile and performed a fit of the individual 24 masses for our clusters. We assumed no intrinsic scatter, and as in Sect.~\ref{sec:freeuppM500}, adopted Gaussian priors centred on the dynamical mass estimates from \cite{2025A&A...693A...2S}.

The results for the tested gNFW cases, given in Table~\ref{tab:uppliterature}, are shown in Fig.~\ref{fig:averageM500}. Violin plots illustrate the best-fit $M_{500}$ distributions for the 24 clusters, for fits with $\eta_T$ as a free parameter.
The statistical averages of the fitted masses over our sample are shown with circles. Diamonds give the average mass for fits with $\eta_T$ fixed to 1.05. In the same figure, we present the distribution and average of the best-fit masses from our joint fits with the universal pressure profile obtained in Sect.~\ref{sec:freeuppM500}. Unsurprisingly, the joint fit of the UPP parameters and individual cluster masses performs the best (in particular, for $\delta=2/3$ and $\delta$ free cases) in modelling the data pressure profiles.

The UPPs we have tested have all been derived from different samples, with individual cluster pressure profiles normalised with masses obtained under different hypotheses. In N07 and A10, the authors used X-ray-only pressure profiles (together with simulations), while PACT and MP23 works are based on tSZ-only analyses. The rest of the studies (P13, B17, G19, S23, and this work) combine X-ray and tSZ data. No differentiation between tSZ and X-ray-based results is observed in Fig.~\ref{fig:averageM500}. Similarly, from the variety of mass and redshift ranges of the samples used in the considered works, we do not observe any obvious trend that could have pointed towards an evolution of the universal pressure profile. In A10, P13, B17, and PACT the provided UPPs were derived by normalising individual pressure profiles with masses estimated from
tSZ or X-ray scaling relations. In G19, authors made use of masses measured from their own HSE mass profiles (based on \xmm\ and \planck\ data), while in MP23 authors employed the SPT masses \citep[][]{2015ApJS..216...27B}, renormalised for compatibility with the A10 mass definition, and N07 assumed \textit{Chandra} HSE masses. Thus, all of the mentioned UPPs were reconstructed by normalising the individual pressure profiles with tSZ or X-ray-based masses, which are known to be biased low as they do not account for the presence of non-thermal pressure in clusters \citep{pratt2019}. Contrarily, in S23 cluster masses were obtained from an X-ray scaling relation calibrated using weak lensing masses \citep{2016MNRAS.463.3582M}.

The masses we derived over our sample of 24 clusters, by assuming the shape of each aforementioned UPP from the literature, are also low against these obtained from our joint fit with the UPP and scaling law parameters (green, red, and blue results in Fig.~\ref{fig:averageM500}). The only exception is N07, who used a sample of relaxed and hot systems ($T_\mathrm{X} > 5$ keV), leading to a peaked and steeper UPP shape (see Fig.~\ref{fig:uppliterature}), and subsequently biasing the associated fitted masses towards higher values. When comparing to N07 (and S23) results one must also bear in mind the temperature cross-calibration difference between \textit{Chandra} and \xmm\ \citep{2015A&A...575A..30S}. 

We note that when fixing the UPP to the A10 profile, we consistently fall back on the MMF3 masses ($< M_{500}^{\mathrm{fit\; A10}}/M_{500}^{\mathrm{MMF3}} > = 0.997 \pm 0.096$), even when taking dynamical masses as priors. MMF3 masses were obtained by measuring the tSZ signal on \planck\ maps (Sect.~\ref{sec:m500s}) assuming the A10 parameters \citep{2014A&A...571A..29P}. This is a further coherence-check demonstrating the validity of our method.

In the end, the average relative difference across the cases we tested is $-23, -23$, and $-24\%$ ($-30, -30$, and $-31\%$ when excluding N07) with respect to our best-fit masses for $\delta=2/3$, $\delta=2/3+0.12$, and $\delta$ free cases, respectively.
The various average cluster mass scales we derived when fixing the UPPs differ by $\sim 10$ to $\sim 50\%$. As we have seen from simulations in Appendix~\ref{sec:simulations}, this stresses the intricate dependence between the mass scale and the universal pressure profile.
Thus, the consistent use of the cluster mass scale, $P_{500}-M_{500}$ scaling relation, and universal pressure profile shape are needed to prevent biases in dependent studies, such as the cosmological analysis based on the tSZ angular power spectrum \citep{2017MNRAS.469..394H,2018A&A...614A..13S,2018MNRAS.477.4957B}. In the reconstruction of the UPP from a given sample, the closer the normalising masses to the true mass scale, the lower the biases on the reconstructed UPP.

Other biases, such as selection effects, also have an impact on the inferred UPP shape and amplitude (e.g., A10). The literature UPPs we have tested here have been built upon samples of clusters with different levels of representativity, and not corrected for selection bias effects\footnote{MP23 corrected for the selection bias of SPT clusters in the SPT-SZ \citep{2015ApJS..216...27B} data.}. The same stands for our current sample of 24 clusters. The full CHEX-MATE sample, as an actual representative sample of the galaxy cluster population at low redshift and at high mass, accounting for the aforementioned selection bias, is appropriate to conduct such a consistent study of the universal pressure profile, the $P_{500}-M_{500}$ scaling, and the masses, to determine the ``absolute mass scale'' of the cluster population.

\section{Summary and conclusions}
\label{sec:conclusion}

In view of the strong correlation between the universal pressure profile shape and the assumed cluster mass scale, we have built a framework that enables, for the first time, the joint fit of the UPP and individual cluster masses. The model accounts for the intrinsic scatter of the pressure profile and for a potential systematic difference between thermal pressure profiles reconstructed from X-ray and tSZ data.

We have applied the method to 24 CHEX-MATE clusters for which \xmm\ and \planck\ thermal pressure profiles and dynamical mass estimates are available. Despite the limitations associated with the small sample size considered in this proof-of-concept paper, the sample is large enough to perform a statistical analysis, and by jointly fitting the individual cluster masses and UPP parameters, we have been able to: 1) reconstruct the best gNFW model describing the shape of the pressure distribution for this sample; 2) propagate the uncertainties of individual cluster masses to the UPP; and 3) refine the individual cluster masses into low scatter mass estimates.

Finally, we have quantified the impact of the assumed UPP shape on the cluster masses measured from the thermal pressure profiles. We confirm the need for a coherent modelling of the pressure. The self-consistent UPP and $M_{500}$ framework is specially suitable for tSZ cosmological statistics (e.g., power spectrum, cluster number counts) where mass scaling assumptions propagate into cosmological parameters.

This work establishes the basis for the universal pressure profile analysis that will be carried out with the full CHEX-MATE cluster sample. Further studies will investigate the physical implications outlined in this paper, will account for the relativistic correction to the tSZ signal, and will explore the impact of a consistent modelling of the pressure and a full propagation of the uncertainties on the cosmological results.

\begin{acknowledgements}
Authors acknowledge the support of the French Agence Nationale de la Recherche (ANR), under grant ANR-22-CE31-0010, and the support of the French National Space Agency (CNES). 
This research was supported by the International Space Science Institute (ISSI) in Bern, through ISSI International Team project \#565 ({\it Multi-Wavelength Studies of the Culmination of Structure Formation in the Universe}). 
S.E., M.S., M.R., H.B., F.D.L., and P.M. acknowledge the financial contribution from the contracts Prin-MUR 2022 supported by Next Generation EU (M4.C2.1.1, n.20227RNLY3 {\it The concordance cosmological model: stress-tests with galaxy clusters}). 
M.S. acknowledges the financial contributions from contract INAF mainstream project 1.05.01.86.10, INAF Theory Grant 2023: Gravitational lensing detection of matter distribution at galaxy cluster boundaries and beyond (1.05.23.06.17). 
S.E., M.R. acknowledge the financial contribution from the European Union’s Horizon 2020 Programme under the AHEAD2020 project (grant agreement n. 871158). 
M.D. acknowledges the support of two NASA programs: NASA award 80NSSC19K0116/  SAO SV9-89010 and NASA award 80NSSC22K0476. 
L.L. acknowledges the financial contribution from the INAF grant 1.05.12.04.01. 
B.J.M. acknowledges support from STFC grant ST/V000454/1.
M.D.P. acknowledges financial support from PRIN-MUR grant 20228B938N {\it"Mass and selection biases of galaxy clusters: a multi-probe approach"} funded by the European Union Next generation EU, Mission 4 Component 1 CUP C53D2300092 0006. 
H.S. acknowledges support by NASA Astrophysics Data Analysis Program (ADAP) Grant 80NSSC21K1571. 
H.B., F.D.L., and P.M. acknowledge support by the Fondazione ICSC, Spoke 3 Astrophysics and Cosmos Observations. National Recovery and Resilience Plan (Piano Nazionale di Ripresa e Resilienza, PNRR) Project ID CN\_00000013 "Italian Research Center on  High-Performance Computing, Big Data and Quantum Computing" funded by MUR Missione 4 Componente 2 Investimento 1.4: Potenziamento strutture di ricerca e creazione di "campioni nazionali di R\&S (M4C2-19)" - Next Generation EU (NGEU), by INFN through the InDark initiative. 
This research was supported by Basic Science Research Program through the National Research Foundation of Korea (NRF) funded by the Ministry of Education (2019R1A6A1A10073887). 
M.G. acknowledges support from the ERC Consolidator Grant \textit{BlackHoleWeather} (101086804). 
G.W.P. acknowledges long-term support from CNES, the French Space Agency.
\end{acknowledgements}

\bibliographystyle{aa} 
\bibliography{references_modifs} 

\begin{appendix}

\section{Uncertainties of dynamical mass estimates}
\label{sec:dynmass}

\begin{figure}[h]
    \centering
    \includegraphics[scale=0.38]{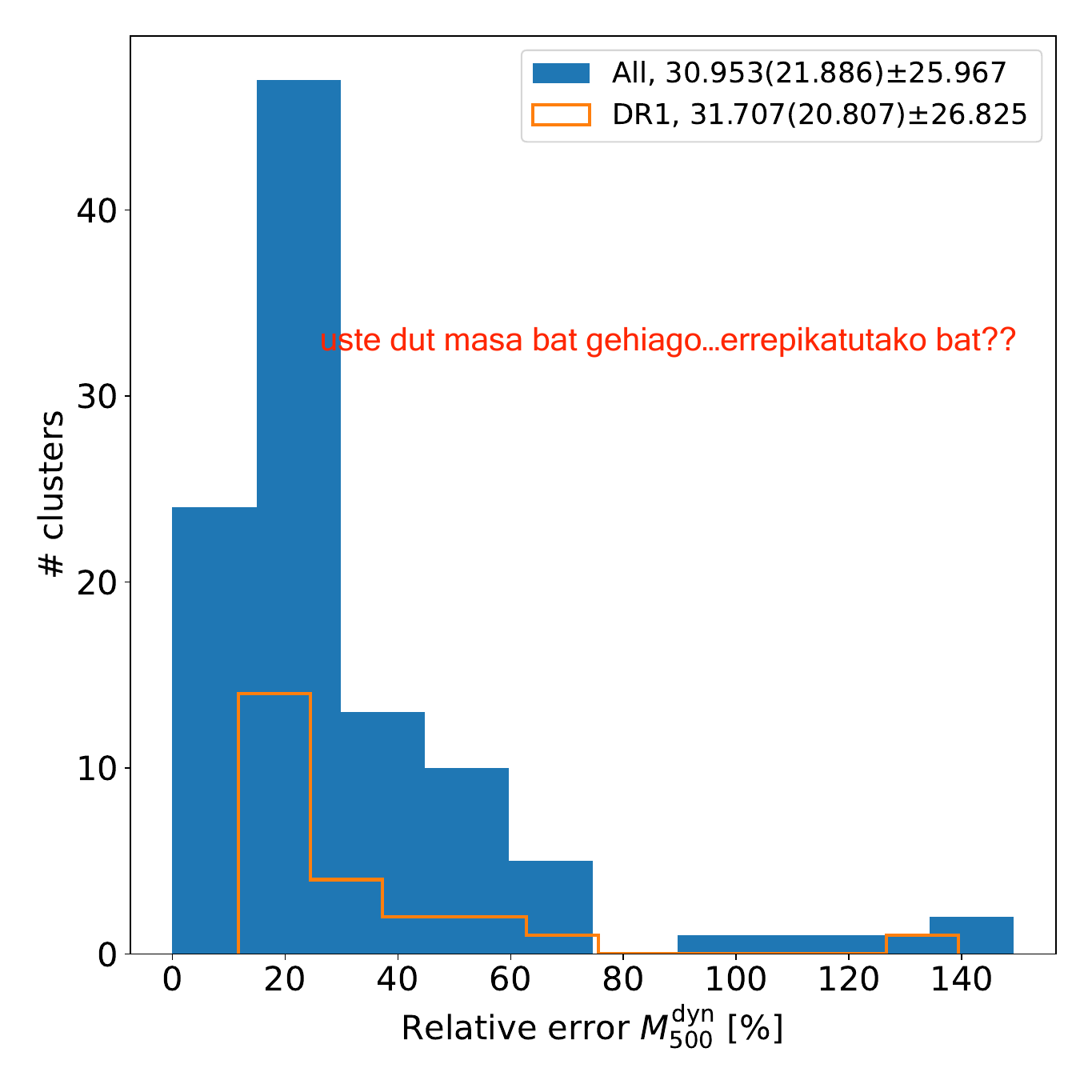}
    \caption{Relative error of individual dynamical mass estimates for the clusters in the CHEX-MATE sample (blue) and the 24 in the DR1 subsample (orange). We give the mean (median) and standard deviation for each distribution.}
    \label{fig:dynerror}
\end{figure}

Fig.~\ref{fig:dynerror} summarises the distribution of the relative uncertainties of dynamical mass estimates for CHEX-MATE clusters as reconstructed in \cite{2025A&A...693A...2S}. In blue, we show the distribution for the clusters in \cite{2025A&A...693A...2S}, while the orange distribution corresponds to the 24 clusters in the DR1 sample. We also give the average relative errors and observe that the DR1 sample is representative of the full CHEX-MATE sample in terms of dynamical mass estimates precision. Thus, we do not expect major changes related to dynamical mass estimates quality when performing the analysis with the full sample.

\section{Intrinsic scatter modelling}
\label{sec:intrinscatter}

\begin{figure*}
    \centering
    \includegraphics[scale=0.33, trim={0cm 0cm 0cm 0cm}]{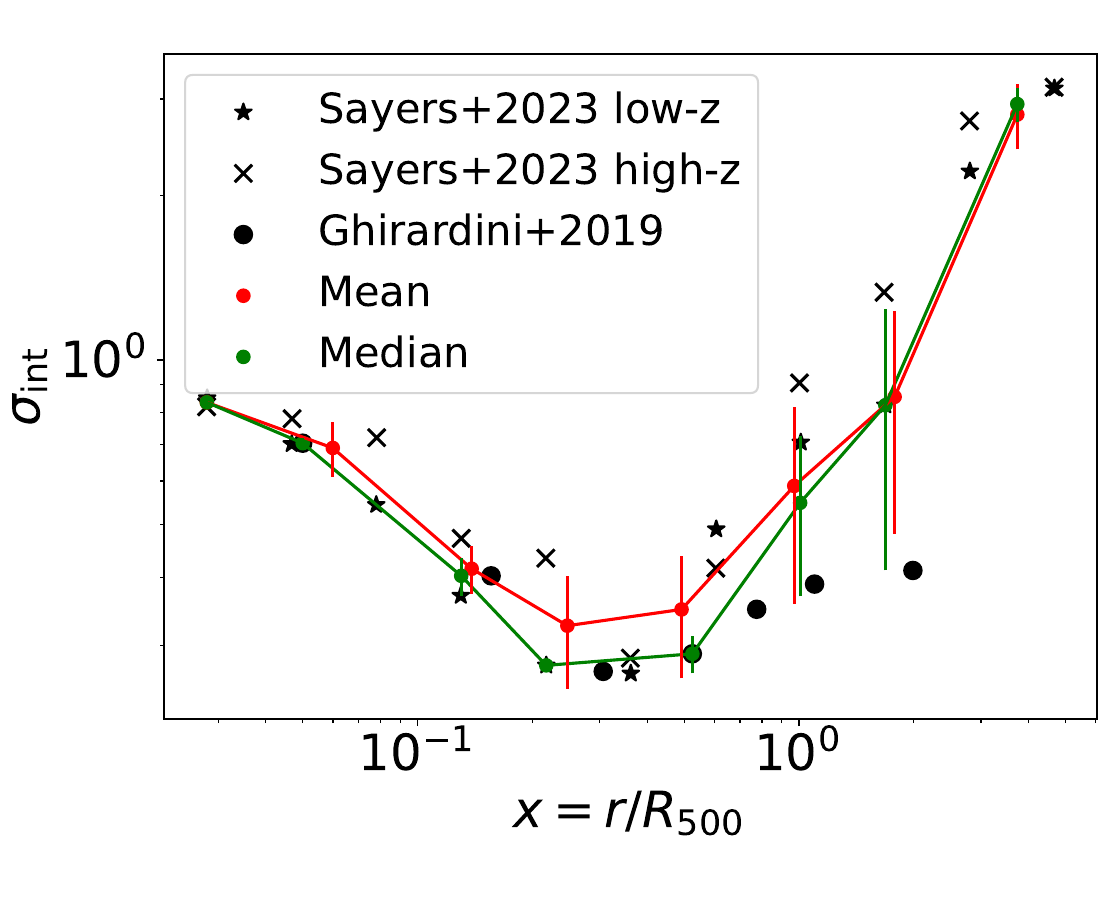}
    \includegraphics[scale=0.33, trim={0cm 0cm 1cm 0cm}]{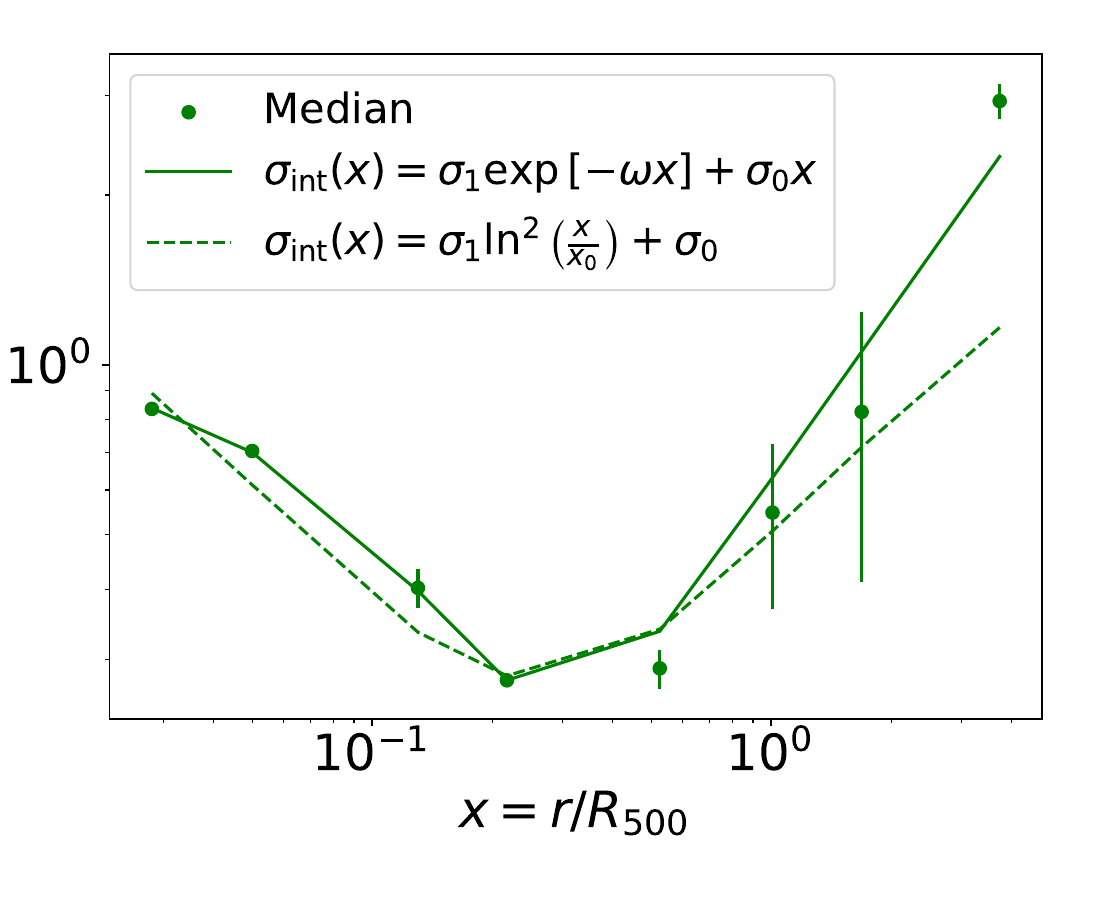}
    \includegraphics[scale=0.33, trim={0cm 0cm 0cm 0cm}]{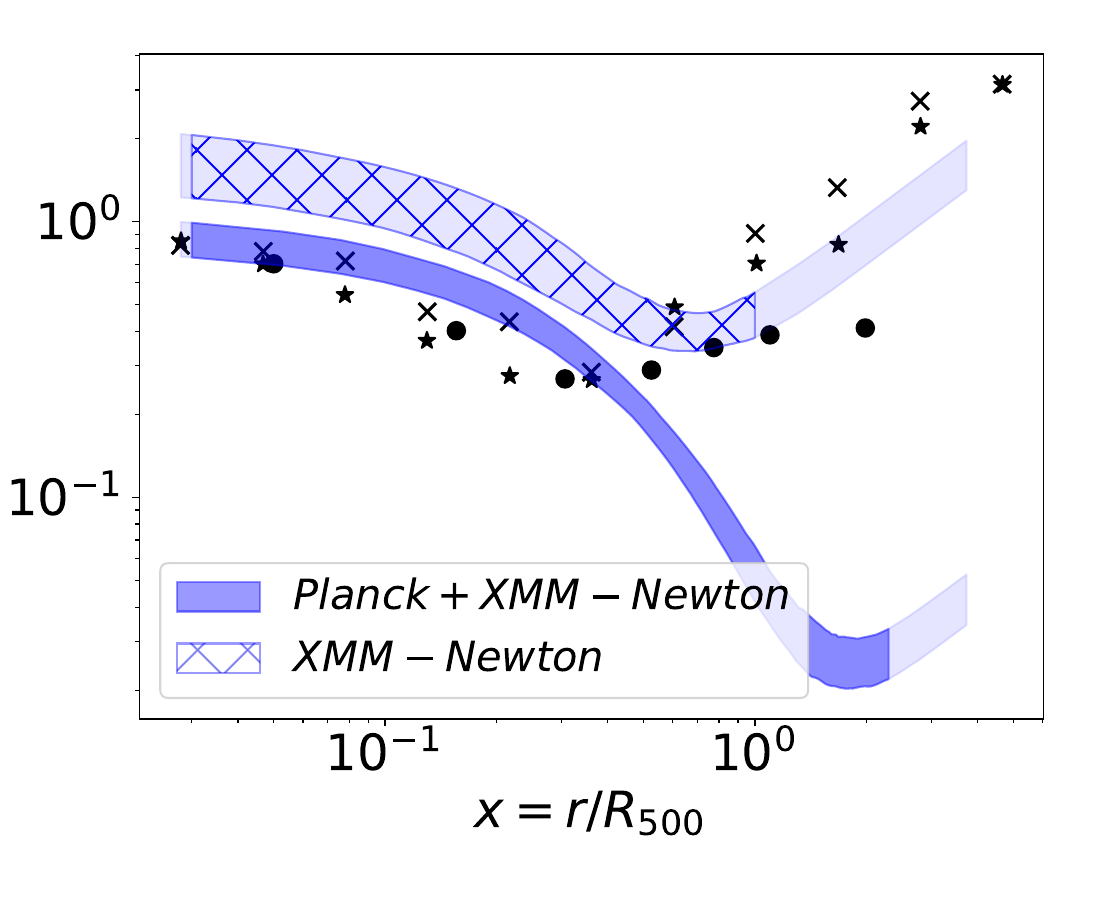}
    \caption{Intrinsic scatter profiles of the thermal pressure distribution in galaxy clusters as a function of normalised radius. Left: black markers show the scatter measured in  \cite{2023ApJ...944..221S} and \cite{2019AA...621A..41G}. Red and green profiles give the mean and median in bins, respectively, with the error bars indicating the standard deviation and the median absolute deviation. Centre: best-fitting scatter models of the median bins following Eq.~\ref{eq:ourscatter} (solid) and Eq.~\ref{eq:ghirardini} (dashed). Right: black markers show the scatter measured in  \cite{2023ApJ...944..221S} and \cite{2019AA...621A..41G} as in the left panel. Blue shaded areas indicate the 16th to 84th percentiles of the intrinsic scatter profile fitted following Eq.~\ref{eq:ourscatter} in the joint fit to our data in Sect.~\ref{sec:freeuppM500}. We show the result from the fit to \planck\ and \xmm\ data as in Fig.~\ref{fig:scatter_fits}, and the hatched area corresponds to the \xmm-only fit. Extrapolated regions for each case are shaded.
    }
    \label{fig:intrinscatter}
\end{figure*}

The intrinsic scatter of the thermal pressure distribution in clusters is expected to decrease from the core out to $\sim 0.5 - 1 \times R_{500}$, where the gravity dominates the structure formation, and then increase towards the outskirts. In Fig.~\ref{fig:intrinscatter}, black markers in the left panel indicate the intrinsic scatter profiles obtained from the joint fit of the UPP and the scatter in \cite{2023ApJ...944..221S} and \cite{2019AA...621A..41G}. We show in red and green their mean and median.

Following the functional form in Eq.~6 in \cite{2019AA...621A..41G}, we can describe the scatter as a function of $\sigma_1, \sigma_0$, and $x_0$:
\begin{equation}
\label{eq:ghirardini}
    \sigma_{\mathrm{int}} (x) = \sigma_1 \ln^2 \left(\frac{x}{x_0} \right) + \sigma_0.
\end{equation}
The dashed line in the central panel in Fig.~\ref{fig:intrinscatter} gives the best-fitting intrinsic scatter profile model describing the median of the data points from the literature with Eq.~\ref{eq:ghirardini}. We observe that the symmetry of this log-parabola shape may force an overestimation (underestimation) of the scatter in the core (outskirts).

As an alternative, we suggest to parametrise the intrinsic scatter profile following Eq.~\ref{eq:ourscatter}, which has the same number of free parameters as Eq.~\ref{eq:ghirardini}. The solid line in the central panel in Fig.~\ref{fig:intrinscatter} shows that Eq.~\ref{eq:ourscatter} enables a better description of the scatter.

In the right panel in Fig.~\ref{fig:intrinscatter} we compare the scatter profiles obtained in \cite{2023ApJ...944..221S} and \cite{2019AA...621A..41G} to the intrinsic scatter we measure in Sect.~\ref{sec:freeuppM500}. The filled area corresponds to the intrinsic scatter profile estimated from the joint fit of \planck\ and \xmm\ pressure bins with $\delta$ as a free parameter. The hatched profile is obtained from the fit of \xmm\ pressure bins only. Extrapolated regions are shaded.

\section{Validation on simulated profiles }
\label{sec:simulations}

The model presented in Sect.~\ref{sec:model} contains a large number of free parameters (global parameters, plus one mass parameter per cluster) and many of them are correlated to each other.
For this reason, it seems necessary to validate the method with end-to-end tests. In this section, we present the construction of simulated mock pressure profiles and the fit of the model to these mocks. We first investigate the ability to recover the input parameters under the impact of different intrinsic scatter levels. Then, we perform the fits by considering the individual cluster masses as free parameters.

\subsection{Construction of simulated profiles}
\label{sec:buildsimulations}

Following the model in Sect.~\ref{sec:model}, we build 
28 pressure profiles corresponding to the DR1 sample by assuming the MMF3 mass for each cluster. We consider $\delta=2/3+0.12$ and $\eta_T=1$, as well as the gNFW parameters from \cite{2010AA...517A..92A}: $[P_0, c_{500}, \alpha, \beta, \gamma] = [8.403, 1.177, 1.0510, 5.4905, 0.3081 ]$. The 28 profiles are drawn from random realisations assuming a given constant intrinsic scatter (not radial dependent), and considering the radial bins corresponding to each data profile (Fig.~\ref{fig:pressureprofs}).

Each simulated profile is associated to the covariance matrix of its corresponding data pressure profile ($\Sigma_{\mathrm{data}}$), including both the \xmm\ and \planck\ parts. By doing this, the fits of simulated profiles in Appendix~\ref{sec:intrinscatterfit} and \ref{sec:simuM500fit} are affected by the same noise level as the data profiles fits in Sect.~\ref{sec:data}. This is crucial to quantify, for instance, the sensitivity to disentangle the various parameters.

\subsection{Sensitivity to intrinsic scatter}
\label{sec:intrinscatterfit}

\begin{figure}[h]
    \centering
    \includegraphics[scale=0.41, trim={1cm 0cm 0 0cm}]{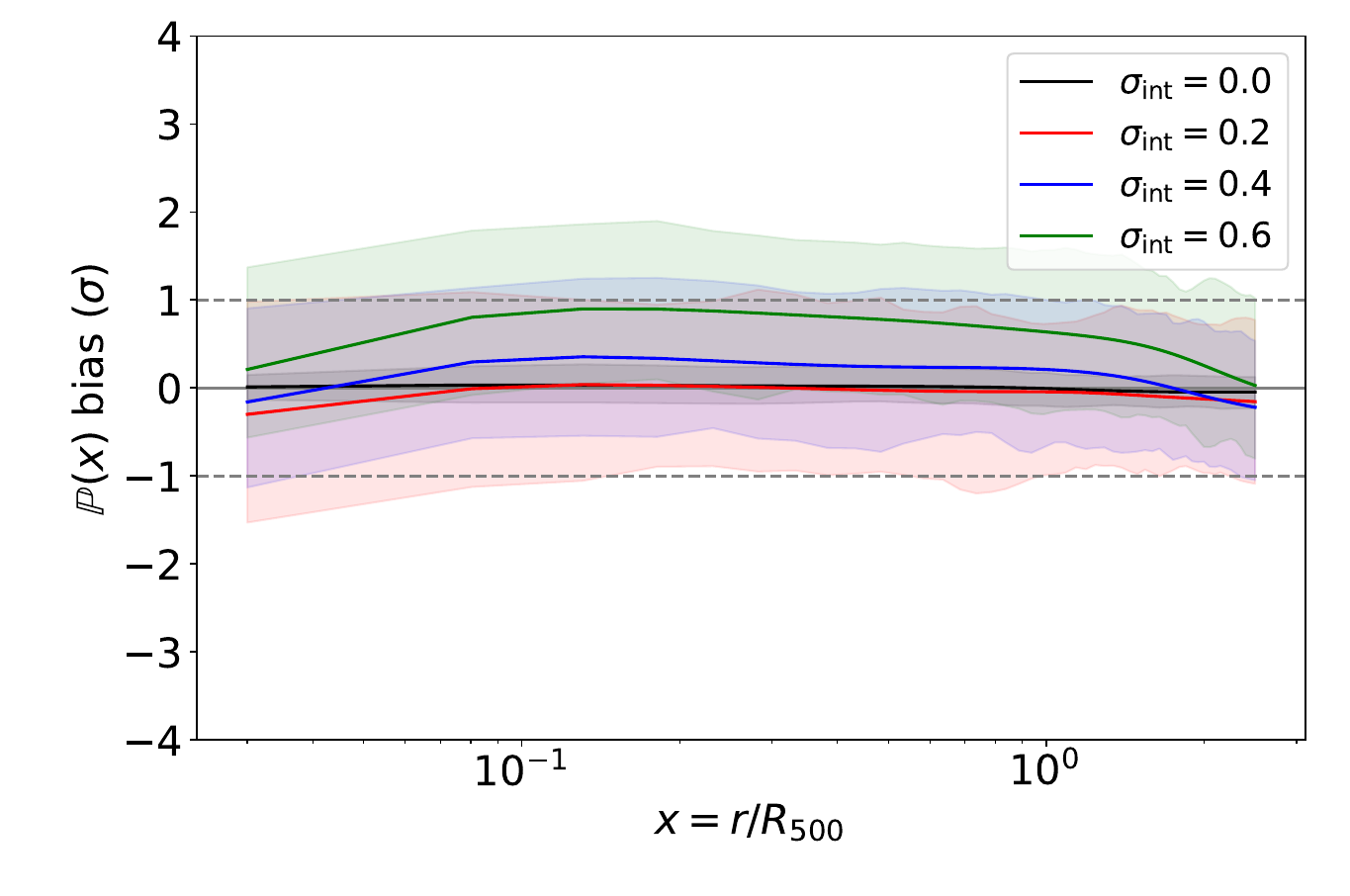}
    \caption{Bias of the universal pressure profile fitted to mock profiles with intrinsic scatter. Colours indicate the results for different input intrinsic scatter values, we only show 0.0, 0.2, 0.4 and 0.6 cases. Solid lines show the mean bias for the 60 realisations and shaded areas cover the 16th to 84th percentiles. Grey dashed lines correspond to $ \pm 1  \sigma$.}
    \label{fig:pressurebias}
\end{figure}
\begin{figure}
    \centering
    \includegraphics[scale=0.0285, trim={5cm 0cm 0 0cm}]{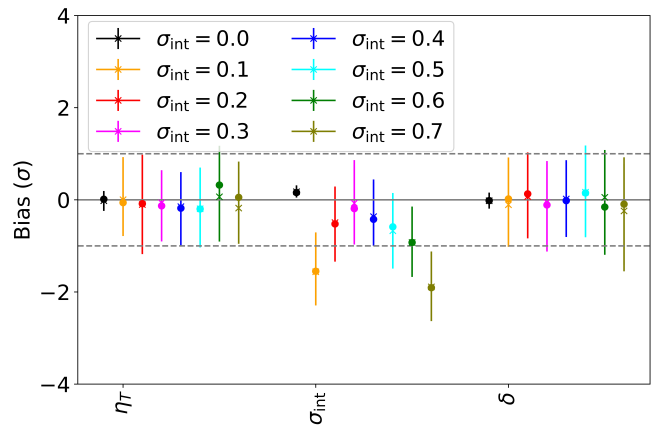}
    \caption{Bias of $\eta_T$, $\sigma_{\mathrm{int}}$, and $\delta$ parameters. Colours indicate the different constant intrinsic scatter levels assumed to create the mock profiles. For each case, the mean (median) bias, and 16th to 84th percentiles are calculated from 60 realisations. 
    }
    \label{fig:nongNFWbias}
\end{figure}

By taking all $M_{500, i}$ masses fixed to the values used to simulate the profiles, in this section we fit all the global parameters, but $L_{\mathrm{int}}$, to profiles generated with different intrinsic scatter levels as input. We test the fitting procedure for 8 constant input scatter values that span a realistic range according to previous works based on observations \citep{2019AA...621A..41G,2023ApJ...944..221S}: $\sigma_{\mathrm{int}}$ in $[0.0, 0.7]$. For each of the input scatter values, we repeat the simulated profiles generation and fitting procedure 60 times to enforce reproducibility. In the fits, we consider uniform priors on the parameters as listed in Table~\ref{tab:parampriors}. We fit a constant intrinsic scatter ($\sigma_{\mathrm{int}} (x) = \sigma_{\mathrm{int}}$) and take the following flat priors: $\sigma_{\mathrm{int}} \sim \mathcal{U}(0, 10)$ and $\delta \sim \mathcal{U}(0, 10)$.

Colour lines in Fig.~\ref{fig:pressurebias} show the bias of the best-fit $ \mathbb{P}(x)$ with respect to the input universal profile when modelling the intrinsic scatter as constant. In particular, they indicate the difference between the best-fit and input $ \mathbb{P}(x)$ in units of the dispersion of the posterior profiles obtained from the fitting procedure. Solid lines and shaded areas represent the mean and 16th to 84th percentiles for 60 realisations, respectively. For the sake of clarity, we do not show the $\sigma_{\mathrm{int}} = 0.1, 0.3, 0.5$, and $0.7$ cases. 

In Fig.~\ref{fig:nongNFWbias}, we present the bias of the best-fit values for the $\eta_T$, $\sigma_{\mathrm{int}}$, and $\delta$ parameters in units of the dispersion of the marginalised posterior distributions obtained from the fitting procedure. We give the mean (median) bias for the 60 realisations with crosses (circles) and error bars indicate the 16th to 84th percentiles. 

We verify (as can be seen in Fig.~\ref{fig:pressurebias}) that we obtain unbiased gNFW profiles if $\sigma_{\mathrm{int}} \le 0.6 $ across the full radial range. When the intrinsic scatter is large, the underlying UPP model is more difficult to recover and up-scattered UPPs are preferred 
(Fig.~\ref{fig:pressurebias}). Given the correlation of the gNFW parameters with $\sigma_{\mathrm{int}}$, the scatter parameter tends towards lower values (Fig.~\ref{fig:nongNFWbias}, and see also Fig.~\ref{fig:scatter_fits} from the fit to data). For intrinsic scatter values that are lower than our average statistical uncertainties ($\sigma_{\mathrm{int}}  \lesssim 0.3 $, see Fig.~\ref{fig:scatter_fits}), it is difficult to distinguish the intrinsic scatter itself from the noise, and the first tends to be underestimated (Fig.~\ref{fig:nongNFWbias}), although within $1\sigma$.
Regarding $\delta$ and $\eta_T$, they are unbiasedly recovered for all the tested intrinsic scatter levels (see Fig.~\ref{fig:nongNFWbias}). Here, we have only validated the 
sensitivity to recover models with constant intrinsic scatter profiles. When letting the $L_{\mathrm{int}}$ parameter free, we test in a single realisation of simulations that the fitted value is compatible with zero. We also model the intrinsic scatter following Eq.~\ref{eq:ourscatter} and obtain an intrinsic scatter profile that decreases with radius (following the tendency observed on data in Sect.~\ref{sec:freeuppM500}) deviating from a constant scatter at a $2\sigma$ level.

\subsection{$M_{500}$ as free parameters}
\label{sec:simuM500fit}

As done for the data analysis in Sect.~\ref{sec:freeuppM500}, in this section, we consider all the individual cluster $M_{500, i}$ masses as free parameters. The correlation between the global parameters ($\{P_0, c_{500}, \alpha, \beta, \gamma; \delta; \eta_T; \sigma_{\mathrm{int}}\}$) and the individual $M_{500, i}$ makes the joint fit very complex. 
Thus, we take Gaussian priors for $M_{500, i}$ (Table~\ref{tab:parampriors}, with the MMF3 mass estimates used to build the mock profiles), assuming the uncertainties on the masses to be of $\sim 25\% $ of the mass values. These uncertainties mimic the typical size of errors associated to dynamical masses (see Sect.~\ref{sec:m500s} and Fig.~\ref{fig:dynerror}).

We perform the fits on mock profiles simulated with $\sigma_{\mathrm{int}} = 0.4$, a reasonable average intrinsic scatter according to observations \citep[see Fig.~8 in][]{2023ApJ...944..221S}. Bias on fits are again quantified as the difference between the best-fit and the input in units of the dispersion $\sigma$ of the posteriors obtained from the fitting procedure: bias = (best-fit $-$ input)/$\sigma$. We test different cases, repeating always the profile creation and fitting procedure 60 times:

\begin{figure}
    \centering
    \includegraphics[scale=0.4, trim={0.5cm, 0cm, 0cm, 0cm}]{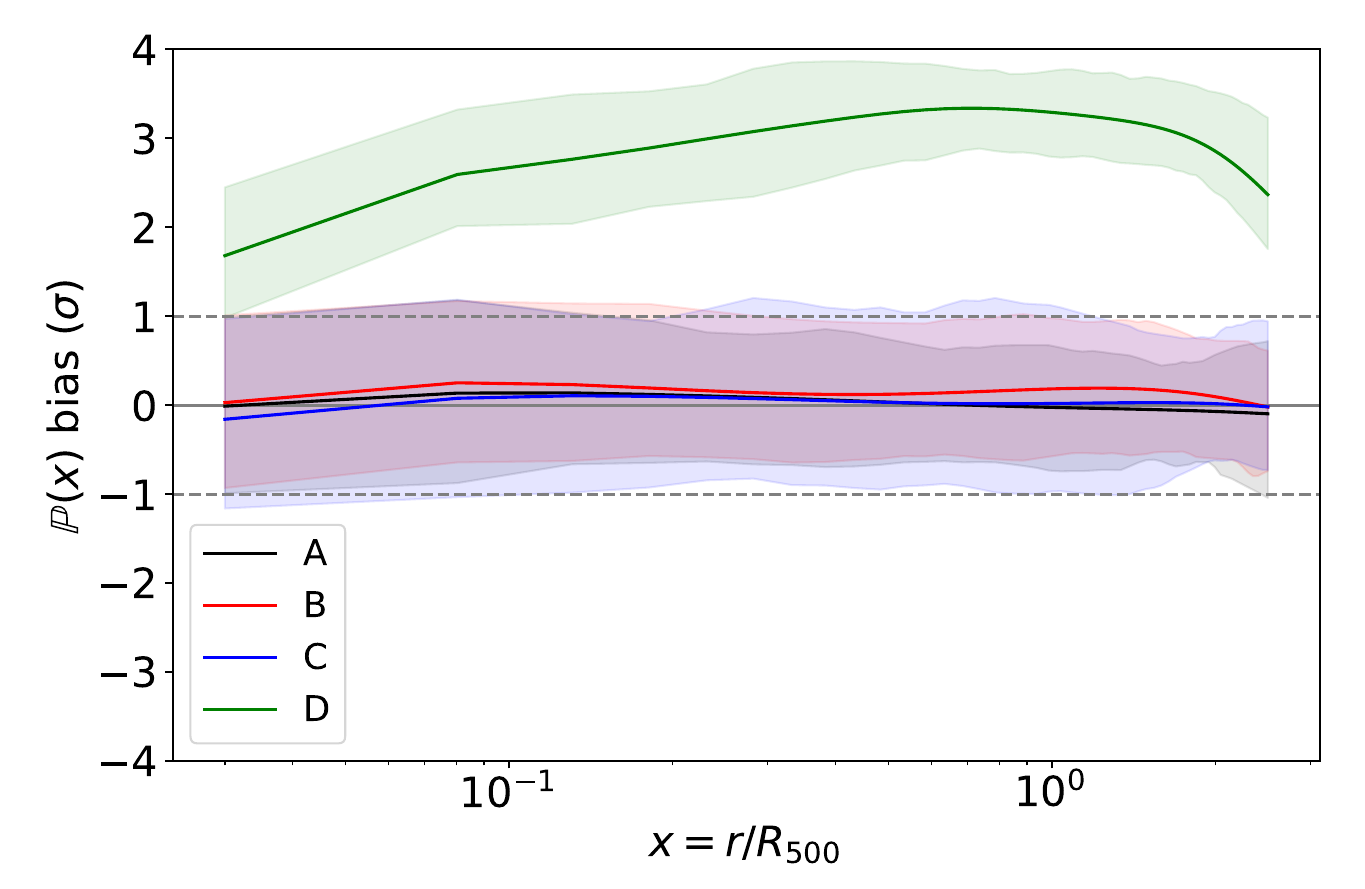}
    \includegraphics[scale=0.405, trim={0.5cm, 0cm, 0cm, 0cm}]{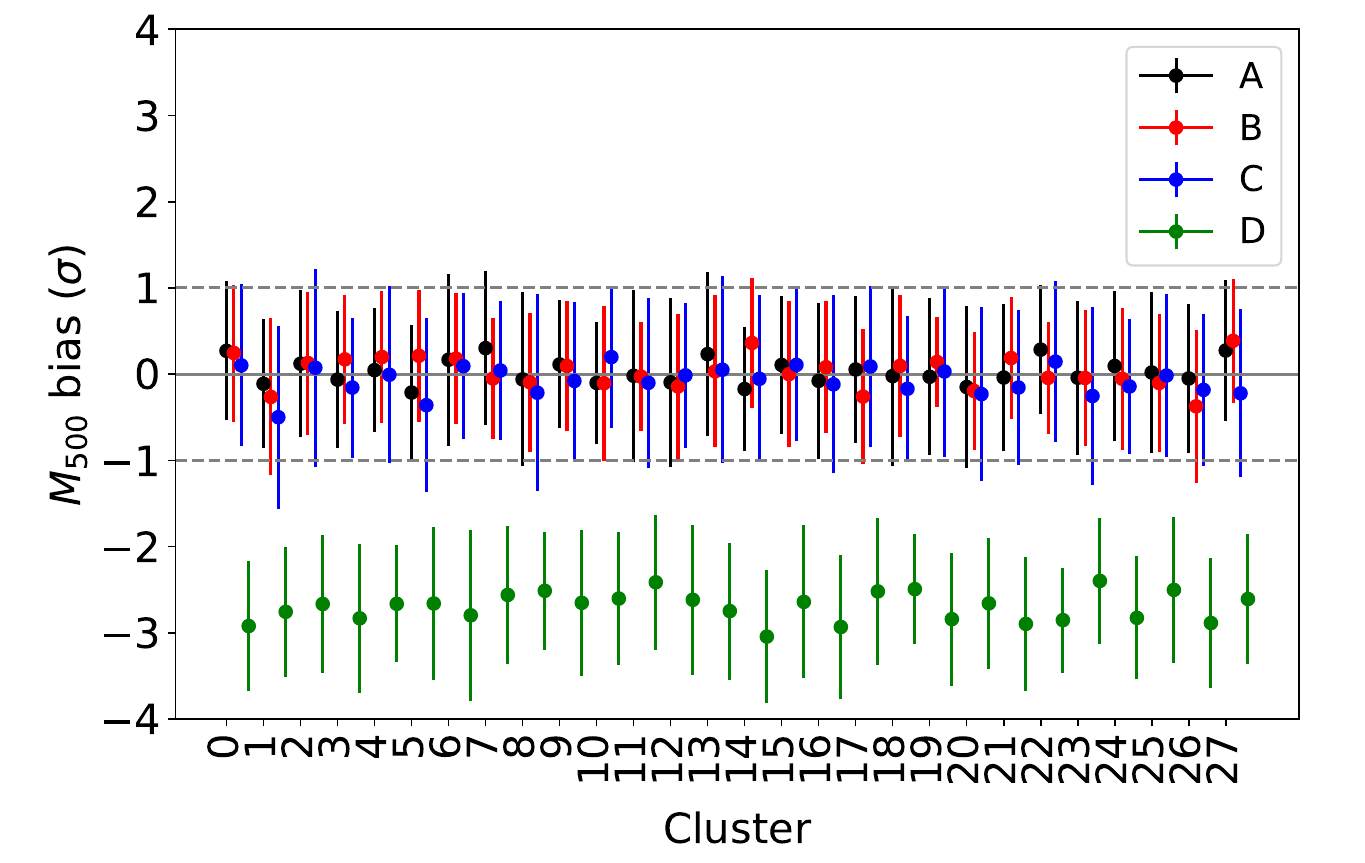}
    \caption{Top: bias of the universal pressure profile fitted to mock profiles with an intrinsic scatter of $\sigma_{\mathrm{int}} = 0.4$. Colours indicate the results with different priors for $M_{500, i}$ and $\delta$, from case A to D, as detailed in the text. Solid lines show the mean bias for the 60 realisations and shaded areas cover the 16th to 84th percentiles. Bottom: bias of the fitted individual cluster masses. For each cluster, we give the median bias and dispersion from the 60 realisations. Dashed lines indicate $ \pm 1  \sigma$.
    }
    \label{fig:massfreefits}
\end{figure} 

\begin{itemize}
    \item Case A: Priors centred on ``true'' $M_{500, i}$. We first consider the case in which the Gaussian priors on $M_{500, i}$ are centred on the mass values used to generate the mock pressure profiles. By fixing the power to $\delta =  2/3+0.12$, the free parameters are $\theta = \{ M_{500, 1}, ..., M_{500, n}; P_0, c_{500}, \alpha, \beta, \gamma; \eta_T; \sigma_{\mathrm{int}} \}$. The black line in the top panel in Fig.~\ref{fig:massfreefits} shows the bias of the fitted UPP with respect to the truth. We observe that the joint fit gives unbiased pressure profiles. On average, fitted masses are completely compatible with input values (bias of $0.02  \pm 0.89 $). Individual biases of the fitted cluster masses are shown in the bottom panel in Fig.~\ref{fig:massfreefits}.
    
    \item Case B: Free $\delta$ and priors centred on ``true'' $M_{500, i}$. Similarly, we try fitting the $P_{500}-M_{500}$ scaling relation together with the rest of the parameters. Since $\delta$ is correlated to $P_0$ and the individual cluster masses, we consider a Gaussian prior for $\delta$. Given the values obtained in the literature (Sect.~\ref{sec:model}), we take a prior for $\delta$ centred on the input value with a scatter of 0.2, that is, $\mathcal{N}(2/3+0.12, 0.2^2)$. We present the bias on the resulting gNFW profile in red in Fig.~\ref{fig:massfreefits}, where we observe that the fitted UPP is unbiased. Compared to Fig.~\ref{fig:pressurebias}, here the bias in units of $\sigma$ is closer to zero because uncertainties derived from posterior distributions are larger, but in both cases the relative bias is of the order of $2.5\%$. In this case, individual masses are also unbiasedly recovered (average bias of $0.04 \pm 0.81$) and the best-fitting $\delta$ are at $-0.47 \pm 0.89$ with respect to the input. Even if within $1\sigma$, this bias on $\delta$ might mean that the mass range covered by the clusters in the sample is not large enough to fully leverage the mass dependence to distinguish between $\delta$ and the rest of free parameters.

    \item Case C: Scattered priors on $M_{500, i}$. When using real data, the mass estimates might be scattered with respect to the true mass of each cluster. In order to simulate this, we repeat the fit but taking mass priors centred on masses scattered by $\sim 20 \%$ with respect to the ``true'' $M_{500, i}$. We consider here fixed $\delta=2/3+0.12$. The blue profile in Fig.~\ref{fig:massfreefits} indicates that even in this case we are able to recover unbiased pressure profiles. Similarly, fitted masses are compatible with ``true'' $M_{500, i}$, the average bias for the masses being of $-0.05 \pm 0.98$.
     
    \item Case D: Biased priors on $M_{500, i}$. Finally, we simulate the case in which the priors on the masses are systematically biased with respect to the ``true'' $M_{500, i}$, that is, the centre of the Gaussian mass priors are shifted with respect to the true masses. By biasing the mass priors low by $\sim 20 \%$, we recover overestimated pressure profiles (green profile in Fig.~\ref{fig:massfreefits}) and underestimated masses (biased by $-2.72 \pm 0.84$). This case shows the impact of mass priors, affecting also the shape of the best-fit gNFW.
    
\end{itemize}

All four cases give unbiased $\eta_T$ reconstructions ($-0.11 \pm 1.02$, $0.12 \pm 1.18$, $0.10 \pm  1.09$, and $-0.19 \pm 0.82$ respectively for A, B, C, and D) and, as in Appendix~\ref{sec:intrinscatterfit}, $\sigma_{\mathrm{int}}$ is systematically underestimated but the average bias is less than the typical measurement uncertainty ($-0.60 \pm 0.94$, $-0.32 \pm 0.87$, $-0.41 \pm 0.90$, and $-0.67 \pm 0.83$ for A, B, C, and D, respectively).

Finally, by fixing the UPP parameters to values different to the ones used to create the mock profiles, we have verified, as expected, that we recover biased cluster masses (average $\sim -4.5\sigma$ and $\sim 1 \sigma$ biases for [$P_0, c_{500}, \alpha, \beta, \gamma$] = [6.72, 1.18, 1.08, 4.30, 0.31] and [$P_0, c_{500}, \alpha, \beta, \gamma$] = [3.36, 1.18, 1.08, 4.30, 0.31], respectively). 
The impact of the assumed gNFW profile on the fitted masses for data pressure profile fits is quantified in Sect.~\ref{sec:fixeduppfits}. We have also verified that we recover a $L_{\mathrm{int}}$ compatible with zero if this is a free parameter in the fit, also in the case in which we fit an intrinsic scatter modelled following Eq.~\ref{eq:ourscatter}. In this case, from the fit of one realisation of simulations we recover an intrinsic scatter profile compatible with a constant at $1\sigma$ level.

\subsection{Summary}

In conclusion, unless the intrinsic scatter of the profiles is larger than $\sigma_{\mathrm{int}} = 0.6 $, 
for fixed $M_{500, i}$ we are able to reconstruct unbiased $\mathbb{P}(x), \delta$, and $\eta_T$. It is the reconstruction of the intrinsic scatter itself that is intricate and will depend on its actual value. It goes without saying that these conclusions also depend on the uncertainties associated to the pressure profile data we are fitting and will change with data quality (Sect.~\ref{sec:data}).

When letting individual cluster masses vary in the fit, we observe that, as expected, the outcomes are biased if mass priors are biased. On the contrary, if our mass estimates are scattered with respect to the true masses, we are able to recover the ``true'' $M_{500, i}$ and unbiased pressure profiles. The free parameters in the model are strongly correlated and the likelihood becomes flat when close to the best-fit values in the multi-variable space. As a consequence, the prior distributions we consider for the mass parameters will determine the scale of the fitted masses.
Regarding the $P_{500}-M_{500}$ scaling relation, it can be difficult to disentangle the power law $\delta$ from the rest of the free parameters in the model, the gNFW amplitude $P_0$ and individual cluster masses being also entangled in the pressure profile amplitude (see Fig.~\ref{fig:cornerall} from the application of the method to data).

\section{Posterior distributions of parameters fitted to data}

\begin{figure*}[h]
    \centering
    \includegraphics[scale=0.265, trim={19cm 0cm 17cm 0cm}]{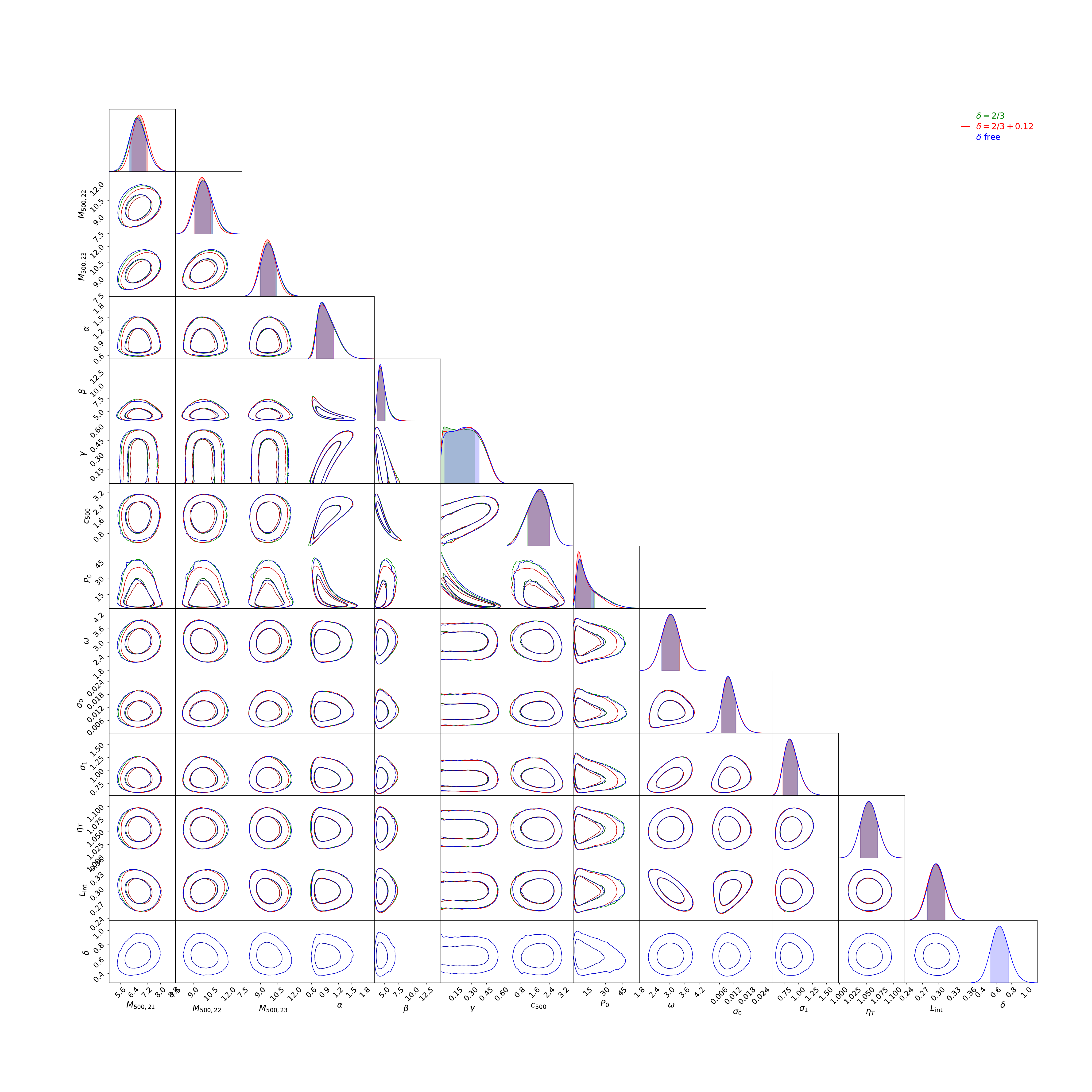}
    \caption{Posterior probability distributions (1D and 2D) of the free parameters fitted to \xmm\ and \planck\ pressure profiles. We only show the masses for 3 of the 24 fitted clusters, illustrating the strong correlations between individual mass parameters. Green, red, and blue correspond respectively to $\delta = 2/3, 2/3+0.12$, and $\delta$ free cases. Cluster masses are given in $10^{14}$~M$_{\odot}$ units.
    }
    \label{fig:cornerall}
\end{figure*}

We present in Fig.~\ref{fig:cornerall} the 1D and 2D posterior distributions for the parameters fitted to data pressure profiles in Sect.~\ref{sec:freeuppM500}.
For visualisation purposes, we only show the posterior probability distributions of three cluster mass parameters as an example to illustrate their degeneracies. Different colours correspond to results for different values of $\delta$.

\begin{figure*}[h]
    \centering
    \includegraphics[scale=0.47, trim={0cm 0cm 0cm 0cm}]{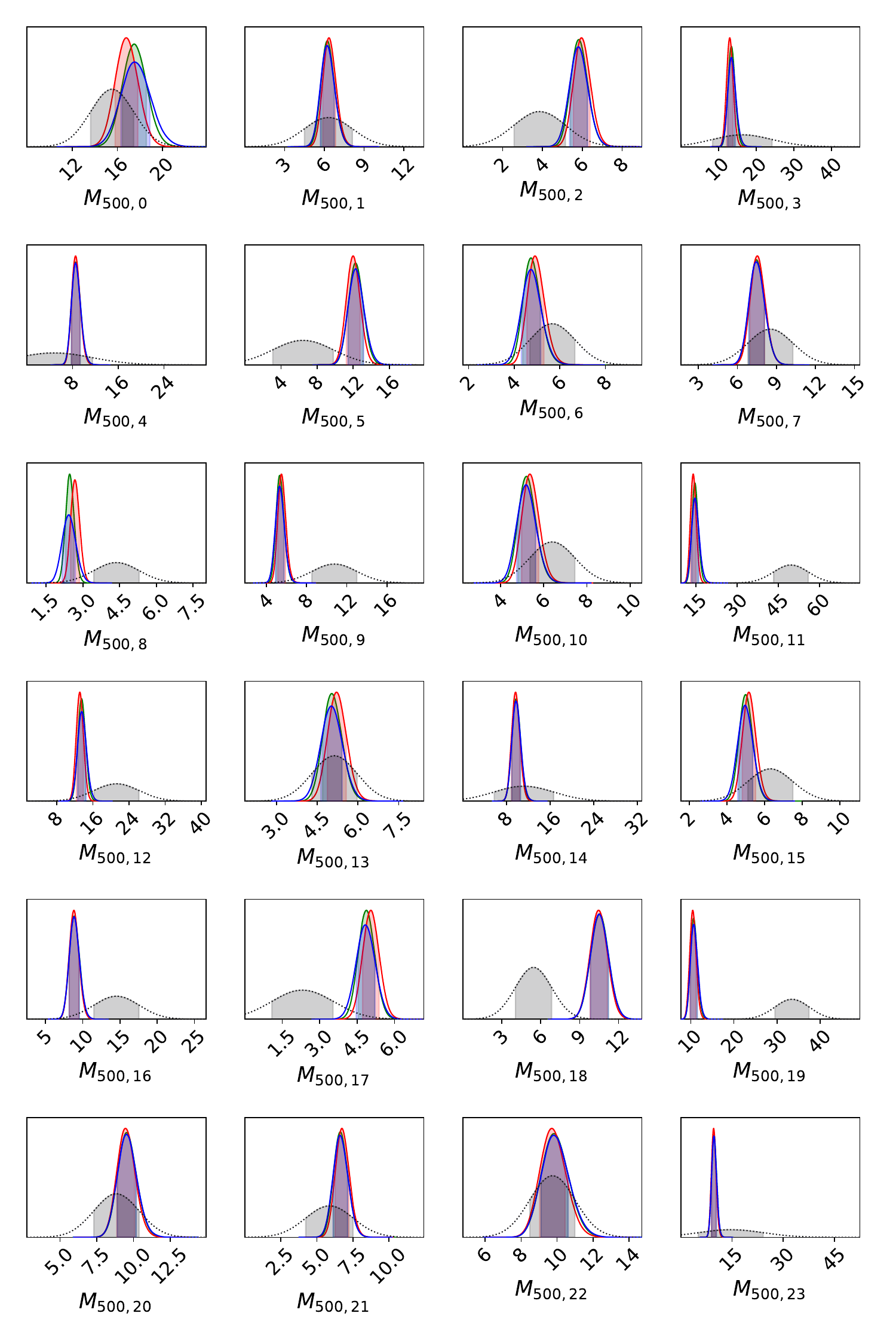}
    \caption{Marginalised posterior distributions of the 24 cluster masses obtained from the joint fits of UPP and $M_{500, i}$ parameters to data pressure profiles. We follow the same color scheme as in Fig.~\ref{fig:cornerall}. In grey we indicate the prior distribution considered for each parameter, corresponding to Gaussian distributions centred on the dynamical mass estimates from \cite{2025A&A...693A...2S} with dynamical mass measurement errors as their standard deviations. All masses are in $10^{14}$~M$_{\odot}$ units. 
    }
    \label{fig:cornermass}
\end{figure*}

In Fig.~\ref{fig:cornermass}, we compare the marginalised posterior distributions of the individual $M_{500, i}$ parameters to the prior distributions assumed for each of them. 


\section{UPP parameters from the literature}
\label{sec:literatureupps}

\begin{figure*}[h]
    \centering
    \includegraphics[scale=0.45]{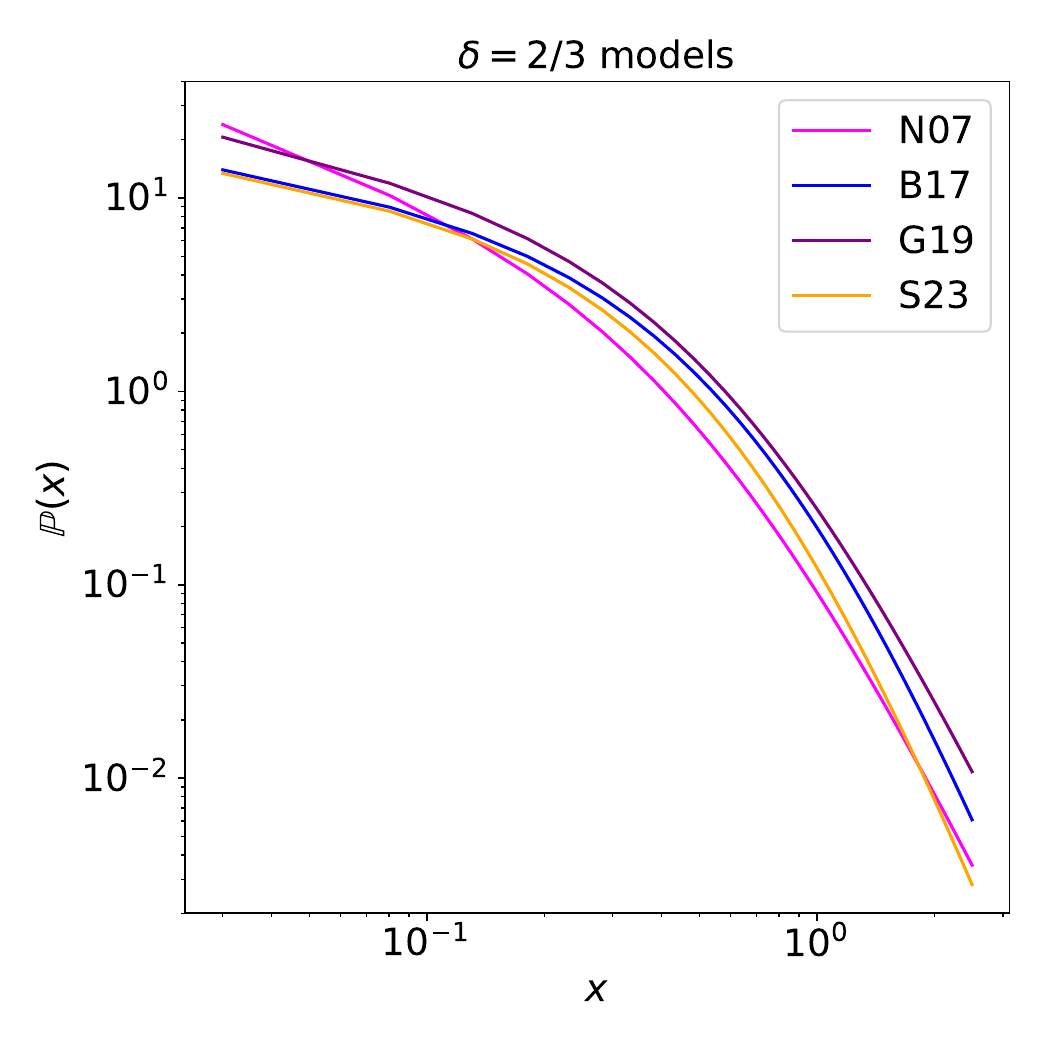}
    \includegraphics[scale=0.45]{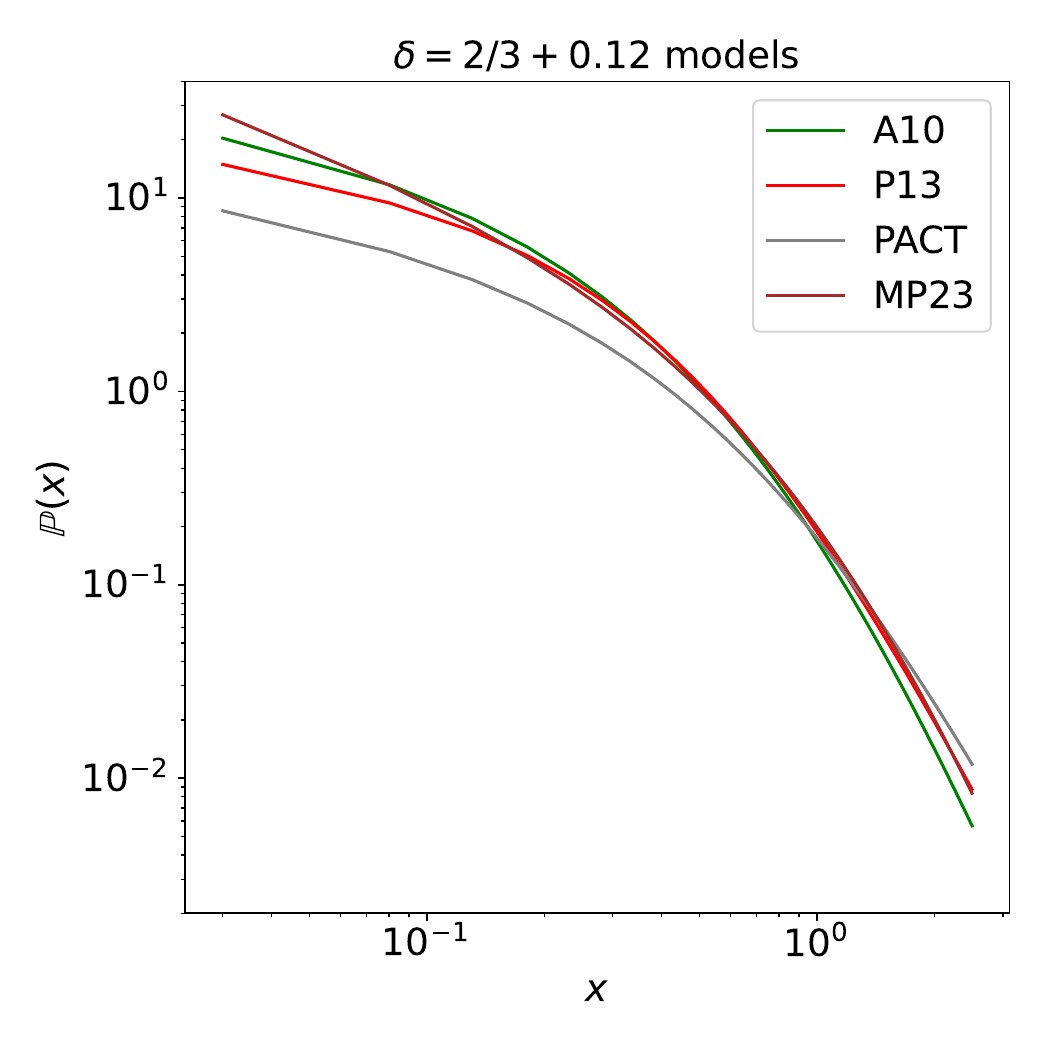}
    \caption{Universal pressure profiles from the literature. The parameters of the gNFW model for each case are summarised in Table~\ref{tab:uppliterature}. We distinguish between the models reconstructed assuming $\delta=2/3$ (left) and $\delta=2/3+0.12$ (right). }
    \label{fig:uppliterature}
\end{figure*}

Table~\ref{tab:uppliterature} summarises the gNFW and $\delta$ parameters from the literature considered in Sect.~\ref{sec:fixeduppfits}. These gNFW profiles are shown in Fig.~\ref{fig:uppliterature}, distinguishing between the models reconstructed assuming $\delta=2/3$ and $\delta=2/3+0.12$.

\renewcommand{\arraystretch}{1.4}   
\begin{table*}[h!]
    \caption{UPP parameters from the literature.}
    \centering
    \begin{tabular}{c c c c c c c c}
    \hline\hline
        Name  &  Reference &  $P_0$  & $c_{500}$ & $\alpha$ & $\beta$ & $\gamma$ & $\delta$ \\\hline
        N07 & \cite{2007ApJ...668....1N} & 3.3 & 1.8 & 1.3 & 4.3 & 0.7 & 2/3  \\  
        A10 & \cite{2010AA...517A..92A} & 8.403 & 1.177 & 1.0510 & 5.4905 & 0.3081 & $2/3+0.12$ \\   
        P13 & \cite{2013AA...550A.131P} & 6.41 & 1.81 &  1.33 & 4.13 & 0.31 & $2/3+0.12$ \\ 
        B17 & \cite{2017ApJ...843...72B}, $z<0.5$ sample & 5.25 & 1.18 & 1.27 &  5.41 & 0.31 & $2/3$ \\ 
        G19 & \cite{2019AA...621A..41G} & 5.68 & 1.49 & 1.33 & 4.40 & 0.43 & 2/3\\ 
        PACT & \cite{2021AA...651A..73P} & 3.36 & 1.18 & 1.08 & 4.30 & 0.31 & $2/3+0.12$\\ 
        MP23 & \cite{2023AA...678A.197M} & 1.70 & 0.61 & 1.05 & 6.32 & 0.71 & $2/3+0.12$\\ 
        S23$^{(*)}$ & \cite{2023ApJ...944..221S} & $10^{0.74}$ & $1.4$ & $10^{0.12}$ & $10^{0.74}$ & 0.3 & $2/3$  \\ \hline
    \end{tabular}
    \tablefoot{We give the values for gNFW parameters and $\delta$ obtained and considered in each work. $^{(*)}$For \cite{2023ApJ...944..221S}, we take $\delta=2/3$ even if deviations to this value were measured at low significance in the work. In addition, we consider that the measured mass and redshift evolutions of the gNFW parameters are not significant and take $a_z=0$ and $a_m=0$ for Eq. 6 in \cite{2023ApJ...944..221S}.}
    \label{tab:uppliterature}
\end{table*}


\end{appendix}

\end{document}